\documentclass[11pt,fleqn]{article}
\usepackage{setspace}
\RequirePackage[OT1]{fontenc}
\RequirePackage{amsthm,amsmath}
\RequirePackage[numbers]{natbib}
\RequirePackage[colorlinks,citecolor=blue,urlcolor=blue]{hyperref}
\usepackage{pgf,tikz}\usepackage{mathrsfs}\usetikzlibrary{arrows}
\DeclareMathOperator{\vect}{vec}
\RequirePackage[OT1]{fontenc}
\RequirePackage{amsthm,amsmath}
\usepackage{float}
\usepackage{fullpage}
\usepackage{amssymb}
\usepackage{graphicx}
\usepackage{dsfont}
\makeatletter
\def\captionof#1#2{{\def\@captype{#1}#2}}
\makeatother
\def\1{\mbox{\bf 1}}
\def\R{\mathbb{R}}

\def\N{\mathbb{N}}
\def\P{\mathbb{P}}
\def\E{\mathbb{E}}

\def\F{\mathbb{F}}
\def\R{\mathbb{R}}

\def\Z{\mathbb{Z}}

\def\v{\mbox{Var\,}}

\newcommand{\footremember}[2]{%
    \footnote{#2}
    \newcounter{#1}
    \setcounter{#1}{\value{footnote}}%
}
\newcommand{\footrecall}[1]{%
    \footnotemark[\value{#1}]%
} 
\newcommand{\dpartial}[2]{
 \frac{\partial {#1}}{\partial {#2}}
}

\newcommand{\ddpartial}[3]{
 \frac{\partial^2 {#1}}{\partial {#2} \partial {#3}}
}

\newtheorem{theo}{Theorem}
\newtheorem{lem}{Lemma}
\newtheorem{prop}{Proposition}

\newtheorem{cor}{Corollary}
\newtheorem{Def/Prop}{Definition-Proposition}

\begin{document}
\date{~} 
\author{Zinsou Max Debaly \footremember{affil}{CREST-ENSAI, UMR CNRS 9194, Campus de Ker-Lann, rue Blaise Pascal, BP 37203, 35172 Bruz cedex, France.}\and
Lionel Truquet \footrecall{affil}\footnote{This work was funded by CY Initiative of Excellence
(grant "Investissements d'Avenir" ANR-16-IDEX-0008),
Project "EcoDep" PSI-AAP2020-0000000013.}}

\title{Multivariate time series models for mixed data}
\maketitle

\begin{abstract}
\noindent
We introduce a general approach for modeling the dynamic of multivariate time series when the data are of mixed type (binary/count/continuous).
Our method is quite flexible and conditionally on past values, each coordinate at time $t$ can have a distribution compatible with a standard univariate time series model such as GARCH, ARMA, INGARCH or logistic models whereas past values of the other coordinates play the role of exogenous covariates in the dynamic. 
The simultaneous dependence in the multivariate time series can be modeled with a copula. Additional exogenous covariates are also allowed in the dynamic.
We first study usual stability properties of these models and then show that autoregressive parameters can be consistently estimated equation-by-equation using a pseudo-maximum likelihood method, leading to a fast implementation even when the number of time series is large. Moreover, we prove consistency results when a parametric copula model is fitted to the time series and in the case of Gaussian copulas, we show that the likelihood estimator of the correlation matrix is strongly consistent. We carefully check all our assumptions for two prototypical examples: a GARCH/INGARCH model and logistic/log-linear INGARCH model. Our results are illustrated with numerical experiments as well as two real data sets.
\end{abstract}
\vspace*{1.0cm}
	
	\footnoterule
	\noindent
	{\sl 2010 Mathematics Subject Classification:} Primary 62M10; secondary 60G05, 60G10.\\
	\noindent
	{\sl Keywords and Phrases:}   time series, mixed data, observation-driven models. \\
	
\section{Introduction}
Analyzing multivariate time series is now a common task in many fields. Many applications of multivariate time series historically come from econometrics or finance and many textbooks such as \citep{intro1} or \citep{intro2} now provide an overview of some interesting models in this context. But the development of multivariate time series analysis has been also connected more recently to others important domains such as in biology (\citep{intro3}), ecology (\citep{intro4}) or industrial production (\citep{intro5}) among others.

However, the literature of multivariate time series analysis is much less developed than in the univariate case. For univariate time series, there already exist many interesting dynamic models depending on the nature of the data. Most of the existing works focus on continuous data with the development of ARMA models (\citep{intro1}) or GARCH models in financial econometrics (see \citep{intro6} for an overview). However, many time series are also related to count data (\citep{Fok}) or categorical data (\citep{Fok1}). Count time series data are for instance systematically encountered when analyzing 
the dynamic of transaction numbers in finance or the number of disease cases in epidemiology (\citep{fernandez}) whereas categorical time series have to be analyzed when studying the dynamic of growth/recession period in economics (\citep{kauppi}), the dynamic of price changes in finance (\citep{russell}) or DNA sequence analysis (\citep{robin}) among others.    
In contrast, multivariate time series are mainly analyzed with continuous models such as vector ARMA models (\citep{intro1}) or multivariate GARCH models (\citep{intro6}). 
It is then difficult to find a flexible approach to obtain multivariate analogues of many univariate models and to then analyze multivariate time series containing either several discrete components or components of mixed type (continuous or discrete). For instance, analyzing the number of transactions in finance of modeling the log-returns of the corresponding asset can only be done separately, though it is quite clear that a bivariate modeling could help to get a better understanding of the mutual interactions of these two quantities across the time.  
Let us mention that a few references are dedicated to multivariate time series for discrete data. For instance, \citep{manner} considered a multivariate binary time series models for analyzing the dynamic of electricity price spikes whereas \citep{DFT} considered recently multivariate time series models for count data such as the numbers of transactions of several assets occurring across the time. The approach used in these two references are (in their spirit) very close to the one we will use in this paper and our general framework contains these two examples as specific cases.   
      
Note that the problem of the modeling of a mixed multivariate response already occurs in the i.i.d. setting and finding suitable multivariate generalized linear models that extend the univariate ones is a non-trivial task. One of the important difficulty is the lack of natural multivariate probability distributions for such data. For i.i.d. data, several approaches have been developed and \citep{de2013analysis} gives an interesting survey of some of them.
However, the treatment of mixed data seems to be devoted to specific cases such as in \citep{YANG} for the joint analysis of continuous/count data.
A notable exception concerns the general regression models considered in \citep{song} with Gaussian copulas and our approach can be seen as a time series analogue of this modeling. Note that our consistency results can also be applied in this context and then provide theoretical guarantees for inference in such models.

In the time series context, our approach consists in using (marginally) some standard univariate time series models called observation-driven. These univariate models are widely popular in the time series literature and provide a sufficiently rich class of dynamics from the GARCH models to Poisson autoregressive INGARCH models (\citep{ferland},\citep{Fok}) or logistic autoregressive processes (\citep{CoxandSnell(1970)}, \citep{Fok1}). Let us also mention that our approach is similar to \citep{Varin}, where a Gaussian copula is used to model panel data of mixed type with specified univariate dynamics.
In what follows, we first present these models in the univariate case and then discuss the multivariate extension we will consider in this paper.  

\subsection{Univariate observation-driven models}

Let $P\left(\cdot\vert s\right)$ be a non-degenerated probability distribution on $E$ (typically, $E=\{0,1\}$, $\N$ or $\R$) and depending on a real-valued parameter $s\in G$ (throughout the paper, $G=\R$ or $G=\R_+$), one can define a discrete-time stochastic process $(Y_t)_{t\in\Z}$ with one-point conditional distribution $P$ in the following way.
For a function $g:G\times E\times \R^m$, we assume that for $A\in \mathcal{B}(E)$,
\begin{equation}\label{odm}
\P\left(Y_t\in A\vert \mathcal{F}_{t-1}\right)=P(A\vert \lambda_t),\quad \lambda_t=g\left(\lambda_{t-1},Y_{t-1},X_{t-1}\right),
\end{equation}
where $(X_t)_{t\in\Z}$ is a covariate process taking values in $\R^m$ and for $t\in\Z$, $\mathcal{F}_t=\sigma\left((Y_s,X_s):s\leq t\right)$.

Standard examples of popular time series models of this type are listed below.
\begin{itemize}
\item
If $P\left(A\vert s\right)=\int_A f\left(y-s\right)dy$ with $f$ a probability density on the real line and $g(s,y,x)=a+bs+cy+d'x$, we obtain the dynamic of the following ARMA$(1,1)$ process with exogenous regressors
\begin{equation}\label{superARMA}
Y_t=a+(b+c)Y_{t-1}+d'X_{t-1}+\varepsilon_t-b \varepsilon_{t-1},
\end{equation}
where $\left(\varepsilon_t\right)_{t\in\Z}$ is a sequence of i.i.d. random variables with probability density $f$. 
\item
If $P\left(A\vert \lambda\right)=\int_A \frac{1}{\lambda}f\left(\frac{y}{\lambda}\right)dy$, we obtain a GARCH type model with volatility process $\left(\lambda_t\right)_{t\in\Z}$ and noise density $f$. Such dynamic is generally
represented in more compact form in the literature,
$Y_t=\varepsilon_t\lambda_t$ where $\left(\varepsilon_t\right)_{t\in\Z}$ is a sequence of i.i.d. random variables with probability density $f$.
\item
For count time series, a standard choice for $p(\cdot\vert \lambda)$ is the Poisson distribution with parameter $\lambda$. We obtain the well-known INGARCH processes.
\item
For binary time series, a classical approach consists in choosing a cdf $F$ on the real line and to set $P(1\vert \lambda)=F(\lambda)$. When $F$ is the cdf of the logistic distribution (resp. Gaussian distribution), we obtain respectively the logistic or probit autoregressive process.
\end{itemize}
Some parametric or semiparametric time series models satisfying (\ref{odm}) are obtained when $g=g_{\theta}$ depends on a finite-dimensional unknown vector of parameters $\theta$. Most of the observation-driven models given above have been studied without exogenous covariates. But some recent theoretical guarantees 
for inclusion of exogenous covariates in non-linear time series models including those mentioned above have been obtained recently (\citep{FT1},\citep{Tru},\citep{DNT},\citep{DT}). 

\subsection{Extension to multivariate mixed time series models}
Our aim is to consider multivariate time series models of type (\ref{odm}). Since there is no natural multivariate distribution $P$ for considering mixed data, a possible approach is to consider multivariate distributions on some Cartesian products $E_1\times\cdots\times E_k$, $k\geq 1$, denoted by $P\left(\cdot\vert s\right)$, with parameter $s=\left(s_1,\ldots,s_k\right)\in G_1\times\cdots\times G_k$ and with specific univariate marginal distributions $P_i\left(\cdot\vert s_i\right)$ for $1\leq i\leq k$.
A natural construction of this type can be obtained from a copula $C$, i.e. a probability distribution on $[0,1]^k$ with uniform marginals. If for $1\leq i\leq k$, $F_{i,s_i}$ denotes the cdf of the probability distribution $P_i\left(\cdot\vert s_i\right)$ and
$U=(U_1,\ldots,U_k)$ follows the distribution $C$, then the random vector $\left(F_{1,s_1}^{-1}(U_1),\ldots,F_{k,s_k}^{-1}(U_k)\right)$ has marginal distributions $P_1\left(\cdot\vert s_1\right),\ldots,P_k\left(\cdot\vert s_k\right)$.
Since the distribution $C$ can have a general form, we then hope that the multivariate distributions $P\left(\cdot\vert\lambda\right)$ obtained in this way to be quite general. As pointed out in \citep{Genest}, copula for discrete data are not unique and lead to interpretation problems and identification issues. However, copula modeling is still a general and valid approach for modeling many stochastic dependence properties between the coordinates, even if some components are allowed to be discrete. 

We now define multivariate time series models with a conditional distribution $P\left(\cdot\vert \cdot\right)$. To this end, we consider a sequence of i.i.d. random vectors $\left(U_t\right)_{t\in\Z}$ such that $U_1=\left(U_{1,1},\ldots,U_{k,1}\right)$ has a probability distribution denoted by $C$.
We then set
$$Y_t=\left(Y_{1,t},\ldots,Y_{k,t}\right)=\left(F_{1,\lambda_{1,t}}^{-1}\left(U_{1,t}\right),\ldots,F_{k,\lambda_{k,t}}^{-1}\left(U_{k,t}\right)\right)$$
and impose a recursive dynamic on the latent process $\left(\lambda_t\right)_{t\in\Z}$ as in (\ref{odm}).

Finally, we define a semiparametric model
\begin{equation}\label{eq::model1}
Y_{i,t}=F_{i,\lambda_{i,t}}^{-1}\left(U_{i,t}\right),\quad 1\leq i\leq k,\quad
\lambda_t=g_{\theta_0}\left(\lambda_{t-1},Y_{t-1},X_{t-1}\right),
\end{equation}
with $\theta_0\in \Theta\subset \R^Q$. 
For conciseness, (\ref{eq::model1}) will be written $Y_t=F_{\lambda_t}^{-1}\left(U_t\right)$.

The paper is organized as follows. In Section \ref{autoreg}, we study stationarity properties of model (\ref{eq::model1}) as well as one particular case with a linear autoregressive function $g$. 
In Section \ref{inference}, we first study inference estimation of parameter $\theta_0$ in the general semiparametric model (\ref{eq::model1}). Here, we will proceed by pseudo-maximum likelihood estimation and it is the copula $C$ can be quite general and the marginal c.d.f. are not necessarily specified, as in the case of ARMA or GARCH components. We next consider a parametric model, with a copula $C$ depending on a finite number of parameters. Here, the marginals c.d.f. will be specified. In this setting we assume that
\begin{equation}\label{model2}
C(du_1,\ldots,du_k)=C_{R_0}(du_1,\ldots,du_k):=c_{R_0}(u_1,\ldots,u_k)du_1\cdots du_k,\quad R_0\in \Gamma\subset \R^{S}.
\end{equation}
For the model defined by (\ref{eq::model1}) and (\ref{model2}), the parameter of interest is the couple $\left(\theta_0',R_0'\right)'$.
We then prove that the conditional distribution $P(\cdot\vert s)=P_{R_0}(\cdot\vert s)$ can be consistently estimated by maximizing the likelihood function. For a Gaussian copula model, we show the the MLE of the correlation matrix $R_0$ is strongly consistent. 
Throughout the paper, we illustrate our results with a bivariate GARCH/INGARCH model for continuous/count time series data and a bivariate logistic/INGARCH model for binary/count time series data.
Numerical experiments and an application of our results to two real data sets is given in Section \ref{numeric} whereas the proofs of all our results are postponed to Section \ref{proof}. Finally an Appendix section gives some auxiliary lemmas needed for the proofs as well as numerical experiments.

\section{Stability properties}\label{autoreg}

\subsection{Existence of stationary solutions}
We provide below a set of sufficient conditions ensuring existence and uniqueness of a stationary and ergodic solution for the recursions (\ref{eq::model1}).
In what follows, for any positive integer $j$, we denote by $\preccurlyeq$ the classical ordering relation on $\R^j$, i.e. $x\preccurlyeq x'$ if and only if $x_i\leq x_i'$ for $i=1,\ldots,j$. Moreover $\vert\cdot\vert_1$ denotes the $\ell_1$ norm on $\R^j$, i.e. $\vert x\vert_1=\sum_{i=1}^j\vert x_i\vert$ for $x\in\R^j$. Moreover, for any matrix $C$, we denote by $\vert C\vert_{vec}$ the matrix of the same size, obtained by replacing the entries of $C$ by their absolute values. Finally, for $t\in\Z$, let $\mathcal{F}_t$ be the sigma-field generated by the random vectors $(U_s,X_s)$, $s\leq t$.

	\begin{description}
		\item[A1.] The process $\left((U_t, X_t)\right)_{t\in\Z}$
		is stationary, ergodic  and for any $t\in\Z$, $U_t$ is
		 independent from $\mathcal{F}_{t-1}$.
		\item[A2.] There exists $s \in G:=G_1\times\cdots\times G_k$ such that :
		$$
		\E[|g(s, F_s^{-1}(U_0), X_0)|_1] < \infty
		$$
		\item[A3.]  There exists a square matrix $H$ of size $k$, with nonnegative elements and such that $\rho(H) < 1$ and a.s.
		$$
	 \forall (s_1,s_2) \in G^2, ~~	\E[|g\left(s_1, F_{s_1}^{-1}(U_0), X_0\right) - g\left(s_{2}, F_{s_2}^{-1}(U_0), X_0\right)|_{vec}\vert \mathcal{F}_{-1}]  \preccurlyeq  H |s_1-s_2|_{vec}
		$$
	\end{description}

\begin{theo}
\label{th::stabilityLatent}
Let Assumptions {\bf A1-A3} hold true. 
There then exists a unique stochastic process $\left(\left(Y_t,\lambda_t\right)\right)_{t\in\Z}$ solution of (\ref{eq::model1}) with $g_{\theta_0}=g$ and which is stationary, $\left(\mathcal{F}_t\right)_{t\in\Z}-$adapted and such that $\E\left(\vert\lambda_0\vert_1\right)<\infty$. Moreover the process $\left((Y_t,\lambda_t,X_t)\right)_{t\in\Z}$ is stationary and ergodic.
\end{theo}

Note that the previous result only guarantees existence of an integrable solution. Sometimes higher-order moment conditions for this solution are required. In the Appendix Section \ref{appendix}, Lemma \ref{lem::momFonc}
gives a useful criterion to check existence of such moments for this unique solution.

\subsection{Specific results for linear type dynamics}
In this section, we consider that for $1\leq i\leq k$, $G_i=G$ where $G$ is either equal to $\R_+$ or to $\R$. we specify the previous results when the latent process follows the dynamic
\begin{equation}
\label{eq::linearla}
   ~  \lambda_t= d + B \lambda_{t-1} + A \overline{Y}_{t-1} + \Gamma X_{t-1}, ~t\in \Z \text{ and }  \theta \in \Theta,   
\end{equation}
with $d \in G^k$, $A$ and $B$ are square matrices of size $k$ and with coefficients in $G$, $\Gamma$ is a matrix of size $k\times m$ and with coefficients in $G$ and $\overline{Y}_{i,t}=g_i\left(Y_{i,t}\right)$ where for $1\leq i\leq k$, $g_i:E_i\mapsto G$ is a measurable mapping.
Though the practical implementation of our models with the dynamic (\ref{eq::linearla}) will be only considered when the matrix $B$ is diagonal, we give below a set of sufficient conditions ensuring {\bf A2-A3} for general matrices $B$. For a vector $c\in\R^k$, we denote by $diag(c)$ the diagonal matrix of size $k$ with diagonal elements $c_1,\ldots,c_k$.
\begin{description}
		\item[L1.] For $s\in G^k$ and $1\leq i\leq k$, the application $g_i\circ F_{i,s_i}^{-1}$ is integrable with respect to the Lebesgue measure on $[0,1]$ and $X_1$ is integrable.
		\item[L2.] For any $1\leq i\leq k$, there exists $c_i>0$ such that for every $(s_i,s_i')\in G^2$,
		$$\int_0^1 \left\vert g_i\circ F^{-1}_{i,s_i}(u)-g_i\circ F^{-1}_{i,s'_i}(u)\right\vert du\leq c_i\left\vert s_i-s_i'\right\vert.$$
		\item[L3.]
		The spectral radius $\rho\left(\vert A\vert_{vec}diag(c)+\vert B\vert_{vec}\right)$ is less than one.
			\end{description}
The following result is a straightforward corollary of Theorem \ref{th::stabilityLatent}. In particular the matrix $H$ in {\bf A3} is given by $\vert A\vert_{vec}diag(c)+\vert B\vert_{vec}$.

\begin{cor}\label{cor::stabilityLinear}
Let Assumptions {\bf L1-L3} and Assumption {\bf A1} hold true. The conclusions of Theorem \ref{th::stabilityLatent} are then valid. 
\end{cor}

\subsection{Examples of linear dynamics}\label{discussion}

For defining multivariate stationary time series models of type (\ref{eq::linearla}), the most constraining assumption to check is Assumption {\bf L2} which imposes, coordinatewise, a Lipschitz type property on the autoregressive function. In the literature, there exist many univariate dynamics satisfying such a property. A general of univariate positive time series models for which such a property holds true has been considered in \citet{davis}, using stochastic ordering properties. In this latter case, $g_i$ is simply the identity function and the distribution $P_i\left(\cdot\vert s_i\right)$, defined from an exponential family, has mean $s_i$. However, there also exist additional dynamics for which {\bf L2} is satisfied and we provide a discussion below.
In what follows, we denote by $(U_1,\ldots,U_k)$ an arbitrary random vector with uniform marginals.
\begin{enumerate}
\item
For count data, a natural univariate dynamic is obtained from the Poisson distribution. A popular one is the linear dynamic, i.e. $g_i(y)=y$ and $F_{i,s_i}$ is the cdf of the Poisson distribution with parameter $s_i>0$. In this case, Assumption {\bf L2} is satisfied with $c_i=1$ from the stochastic ordering property, i.e. $F^{-1}_{i,s_i}\leq F^{-1}_{i,s'_i}$ if $s_i\leq s'_i$ and the fact that $\E F_{i,s_i}^{-1}(U_i)=s_i$. See in particular \citet{davis}, Proposition $4$ and its proof. 
To accommodate with negative correlations, one can define a log-linear model as in \citet{FokTjo}. In this case, we set $g_i(y)=\log(1+y)$ and $F_{i,\lambda_i}$ denotes the Poisson distribution of parameter $\exp\left(s_i\right)$. In this case, {\bf L2} is satisfied. A proof can be found in \citet{FokTjo}, see the proof of their Lemma $2.1$. For the reader convenience, we give a different proof here. From stochastic ordering and the monotone property of the logarithm function, if $s_i\leq s_i'$, we have
\begin{equation}\label{loglinear} 
\E\left\vert \log\left(1+F_{i,s_i}^{-1}(U_i)\right)-\log\left(1+F_{i,s'_i}^{-1}(U_i)\right)\right\vert= \E\log\left(1+F_{i,s'_i}^{-1}(U_i)\right)-\E\log\left(1+F_{i,s_i}^{-1}(U_i)\right)\leq s_i'-s_i.
\end{equation}
The last inequality can be obtained from the mean value theorem, by noticing that if $X_{\mu}$ follows a Poisson distribution of parameter $\exp(\mu)$, then $f:\mu\rightarrow \E\log(1+X_{\mu})$ has a derivative given by 
$$f'(\mu)=\sum_{k\geq 0}\log\left(1+\frac{1}{k+1}\right)e^{-e^{\mu}}\frac{e^{\mu (k+1)}}{k!}\leq 1,$$
if we use the inequality $\log(1+x)\leq x$ for $x\geq 0$. 
\item
Let us next discuss the case of binary time series by assuming that $F^{-1}_{i,s_i}(U_i)=\mathds{1}_{U_i>1-F(s_i)}$ meaning that $P_i(\cdot\vert s_i)$ is the Bernoulli distribution on parameter $F(s_i)$
where $F$ is a given cdf. See \citet{Fok1} for the stability property of such univariate dynamics. We have here
$$\E\left\vert F^{-1}_{i,s'_i}(U_i)-F^{-1}_{i,s_i}(U_i)\right\vert\leq \left\vert F\left(s'_i\right)-F\left(s_i\right)\right\vert\leq c_i\left\vert s_i-s'_i\right\vert.$$
Here $c_i$ denotes the Lipschitz constant of $F$. Two well-known cdf $F$ are widely used in practice, the logistic $F(\mu)=(1+e^{-\mu})^{-1}$, $\mu\in\R$, for which $c_i=1/4$ and the cdf of the standard Gaussian distribution and for which $c_i=1/\sqrt{2\pi}$ (probit model).
\item
Finally, let us discuss the case of continuous components. For a GARCH component, $P_i\left(\cdot\vert s_i\right)$ is the probability distribution of $\sqrt{s_i}\epsilon$ where $\epsilon$ is a centered random variable with unit variance. We then have $F^{-1}_{i,s_i}(U_i)=\sqrt{s_i}F_{\epsilon}^{-1}(U_i)$ where $F_{\epsilon}$ is the cdf of $\epsilon$. If $g_i(y)=y^2$, it is easily seen that {\bf L2} is satisfied with $c_i=1$.
 One can also consider the log-GARCH model which does not impose any positivity condition on lag parameters. Log-GARCH models are discussed in \citep{intro6} and are the analogs of log-linear Poisson autoregressions for count data. With our formulation, $P_i\left(\cdot\vert s_i\right)$ is now the probability distribution of $\exp\left(s_i/2\right)\epsilon$ (a linear dynamic is specified on the logarithm of the conditional variance). Setting $g_i(y)=\log(y^2)$ and assuming that $\P(\epsilon=0)=0$, {\bf L2} is satisfied with $c_i=1$.
  
Another interesting dynamic concerns the linear ARMA$(1,1)$ dynamic. As explained in the introduction, this dynamic is equivalent, up to a reparametrization, to the case where $P_i\left(\cdot\vert s_i\right)$ is the probability distribution of the sum $s_i+\epsilon$, with $\epsilon$ a centered random variable. It is then possible to check {\bf L2} with $g_i(y)=y$ and $c_i=1$. 
\end{enumerate}

\paragraph{Notes}

\begin{enumerate}
\item
For the GARCH models and linear Poisson GARCH models, the latent processes $\lambda_{i,t}$ are required to take positive values as they represent the conditional standard deviation and the intensity respectively. The dynamic parameters are also required to be positive.
One can then combine these two univariate dynamics to construct a bivariate time series model with continuous/count components. This model will be presented in the next section.
Univariate models without any sign restriction on the latent process, such as log-linear Poisson autoregressions, log-GARCH, ARMA and binary time series 
can be used for modeling trivariate time series with continuous/count/binary components. 
Technically, the general model \eqref{odm} can also be used for combining any dynamics of the previous type (whatever the signs of the univariate latent processes). However in this case, specifying a function $g$ preserving the sign constraints could appear to be more arbitrary.

\item
As explained above, a linear type equation (\ref{eq::linearla}) is already interesting for generalizing well-known univariate dynamics and Corollary \ref{cor::stabilityLinear} provides a result for stability for the model. Such a result will be applied to two examples studied in details in the rest of the paper. Its main interest is pedagogical as it illustrates that many classical univariate models can be combined together for defining a multivariate times series model. But we point out that Corollary \ref{cor::stabilityLinear} is not necessarily sharp with respect to Theorem \ref{th::stabilityLatent}. For instance, assume (\ref{eq::linearla}) with {\bf L1-L2} satisfied and with a component, say $i$, defined from the log-GARCH model. In this case, we have
$$\log\left(F_{i,\lambda_i}^{-1}(U_i)^2\right)=\lambda_i+\log\left(F_{\eta}^{-1}(U_i)^2\right)$$
and one can directly check Assumption {\bf A3} with a matrix $H$ such that $H(\ell,i)=\vert A(\ell,i)+B(\ell,i)\vert$ and $H(\ell,j)=\vert A(\ell,j)\vert c_j+\vert B(\ell,i)\vert$ if $j\neq i$. Condition $\rho(H)<1$ is less restrictive than {\bf L3} in this case because of the inequalities
$$H(\ell,i)\leq \left(\vert A\vert_{vec}diag(c)+\vert B\vert_{vec}\right)(\ell,i).$$ 
The same improvement can be obtained if we consider an ARMA component.

\item
Our framework also includes some multivariate time series models for discrete data found in the literature. \citet{manner} considered a multivariate binary time series models with applications to electricity price spikes. The conditional distribution of each marginal can be logistic, Gaussian or of a more general form and the dynamic on the latent process is similar to (\ref{eq::linearla}). A copula structure is also used for modeling the  simultaneous dependence for the multivariate time series. 
In \citet{DFT}, multivariate count autoregressions have been introduced. In these models, the conditional distribution of each marginal is Poisson and both the linear and the log-linear case are studied. The simultaneous dependence is also based on a copula. A main difference with our approach concerns the generations of univariate Poisson marginal distributions. While we use directly the inverse of the Poisson cdf to construct our model, \citet{DFT} simulates several independent copies of the copula to generate exponential inter-arrival times of a Poisson process. However, both models have very similar properties. 
\end{enumerate}

\subsection{Two specific examples}

\subsubsection{The model GAIN}
The GARCH-INGARCH (abbreviated as GAIN) mixed model combines the dynamic of the univariate GARCH model of \citep{Bollerslev(1986)} and the Poisson autoregressive model called INGARCH in \citep{ferland2006integer}. Here, $E_1=\R$, $E_2=\N$, $G_1=G_1=\R_+$ and we define the model as follows.

\begin{equation}
\label{eq::modelGAIN}
    Y_t = \left\{
		\begin{array}{lcl}
		Y_{1,t}  & = &  \lambda_{1,t}^{1/2} F_\epsilon^{-1}(U_{1,t}) \\
		Y_{2,t}  & = &  \inf\left\{y\in\N : \sum_{j = 0}^y e^{-\lambda_{2,t}} \frac{\lambda_{2,t}^j}{j!}  \geq  U_{2,t}\right\}  \\
         \end{array} 
    \right. 
    \end{equation}
    \begin{equation*}
    \lambda_t  = d + B \lambda_{ t-1}  	+ A  \overline{Y}_{t-1}  + \Gamma X_{t-1}
 \end{equation*}
where  $\overline{Y}_{t} = ({Y}_{1,t}^2, {Y}_{2,t})'$ and $F_\epsilon^{-1}$ stands for the inverse of the cumulative probability function of a centered random variable  $\epsilon$ with unit variance. The elements of $d$, $A$ and $B$ are assumed to be nonnegative.
The following result gives a necessary and sufficient condition for the existence of some solutions.

\begin{prop}
\label{pp::stationaryGAIN}
Consider the model \eqref{eq::modelGAIN} and let Assumption {\bf A1} holds true with $X_1$ integrable.
\begin{enumerate}
\item
If $\rho(A+B)<1$, there exists a unique solution $(Y_t)_{t\in\Z}$ to \eqref{eq::modelGAIN} such that $\left((Y_t,\lambda_t)\right)_{t\in\Z}$ is a stationary, $\left(\mathcal{F}_t\right)_{t\in\Z}-$adapted and integrable process. Moreover, the process $\left((Y_t,\lambda_t,X_t)\right)_{t\in\Z}$ is stationary and ergodic.
\item
Conversely, assume that $d$ has positive coordinates and suppose that $\left((Y_t,\lambda_t)\right)_{t\in\Z}$ is a stationary, $\left(\mathcal{F}_t\right)_{t\in\Z}-$adapted and integrable process solution of \eqref{eq::modelGAIN}. Then $\rho(A+B)<1$.
\end{enumerate}
\end{prop}

We next give a result for existence of higher-order moments.

\begin{prop}
\label{pp::momentGAIN}
Consider the model \eqref{eq::modelGAIN} under the assumptions of Proposition \ref{pp::stationaryGAIN} and assume that for some integer $r\geq 1$, $\E(|X_{0}|_ 1^r) < \infty.$ If in addition $\E^{1/r}[\epsilon ^ {2r}]<\infty$ and  
     $\rho\left(B + A \mathrm{diag}(\E^{1/r}[\epsilon ^ {2r}], 1 )\right)< 1$, then
$$
\E(|\overline{Y}_{0}|_ 1^r) < \infty \text{ and }
\E(|\lambda_{0}|_ 1^r)  < \infty.
$$
\end{prop}

\paragraph{Note.}

Under the stationarity condition $\rho(A+B)<1$ and if there exists $r'>1$ such that $\E\epsilon^{2r'}<\infty$ and $\E\vert X_0\vert_1^{r'}<\infty$, one can always find $r>1$ such that $\rho\left(B + A \mathrm{diag}(\E^{1/r}[\epsilon ^ {2r}], 1 )\right) < 1$. 
Existence of a moment of order larger than $1$ is then obtained without any restriction on the lag parameters $A$ and $B$. This property will be particularly important for proving consistency and asymptotic normality of pseudo-likelihood estimators.

\subsubsection{The model BIP}
In what follows, we consider a bivariate time series model compatible with sequences of binary/count data.
This model, called Binary-Poisson (abbreviated as BIP) mixed model, combines an autoregressive logistic model with a log-linear Poisson autoregressive model. 
For the sake of simplicity, we consider a model with two coordinates but extensions including several binary/count time series is straightforward. Here $E_1=\{0,1\}$, $E_2=\N$ and $G_1=G_2=\R$.
The model writes as follows.
\begin{equation}
\label{eq::modelBIP}
    Y_t = \left\{
		\begin{array}{lcl}
		Y_{1,t}  & = &  \mathds{1}_{\{ U_{1,t} \geq 1-F(\lambda_{1,t})\} },\quad F:s\mapsto \frac{1}{1+\exp(-s)} \\
		Y_{2,t}  & = &  \inf\left\{y\in \N: \sum_{j = 0}^y e^{-e^{\lambda_{2,t}}} \frac{e^{j\lambda_{2,t}}}{j!}  \geq  U_{2,t}\right\}  \\
         \end{array} 
    \right. 
    \end{equation}
    \begin{equation*}
    \lambda_t  = d + B \lambda_{ t-1}  	+ A  \overline{Y}_{t-1}  + \Gamma X_{t-1}
 \end{equation*}
where  $\overline{Y}_{t} = ({Y}_{1,t}, \log(1+{Y}_{2,t}))'$. Here the coefficients in $d,A,B,\Gamma$ and the covariate process $(X_t)_{t\in\Z}$ can take arbitrary signs.

\begin{prop}
\label{pp::stationaryBIG}
Consider the model \eqref{eq::modelBIP} and suppose that Assumption {\bf A1} holds true with $X_0$ integrable. Assume furthermore that $\rho(|B|_{vec} + |A|_{vec} \mathrm{diag}(1/4, 1)) < 1$.
Then the conclusions of Corollary \ref{cor::stabilityLinear} are valid and we also have $\E(|\overline{Y}_0|_1) < \infty.$
\end{prop}

\paragraph{Note}

Contrarily to the model GAIN, our conditions for stationarity are not optimal. This is already the case for the univariate log-linear Poisson autoregressive model for which our stability condition is equivalent to $\vert A_{2,2}\vert+\vert B_{2,2}\vert<1$. See \citet{Douc}, Proposition $17$ for a sharper result.
For univariate logistic autoregressions, our condition writes as $\vert A_{1,1}\vert/4+\vert B_{1,1}\vert<1$. This condition is similar to that of \citet{Fok1}
but much more restrictive than the condition $\vert B_{1,1}\vert<1$ given in \citet{Truquet} or in \citet{Truquet2}, Proposition $2$.  
Our results seems to give optimal conditions for some dynamics with positive latent processes, such as for the model GAIN or for multivariate linear Poisson autoregressions. In the latter case, see \citet{DT2}, Theorem $4$, where a similar stability result was applied. 
On the other hand, we point out that our approach can be applied to many nonlinear multivariate dynamics and allows exogenous covariates not necessarily strictly exogenous (i.e. the noise process $(U_t)_{t\in\Z}$ and the covariate process $(X_t)_{t\in\Z}$ are not necessarily independent).

We now investigate existence of some higher-order moments that will be necessary for statistical inference. To this end, we denote by $\overline{A}$ the matrix obtained by replacing the first column of $A$ by the null vector $\textbf{0}$.
We recall that for a matrix $C$ of size $k\times k$, its infinite norm, denoted by $\vert C\vert_{\infty}$, is defined by $\vert C\vert_{\infty}=\max_{1\leq i\leq k}\sum_{j=1}^k\vert C(i,j)\vert$.

\begin{prop}
\label{pp::momentBIG}
Consider the model \eqref{eq::modelBIP} and assume that the assumptions of Proposition \ref{pp::stationaryBIG} are valid. Suppose furthermore that  $| |\overline{A}|_{vec} + |B|_{vec} |_{\infty} < 1$ and that for any $r>0$, $\E\left(\exp\left(r\vert X_0\vert_1\right)\right)<\infty$. 
Then, for any $r>0$, we have
$$\E\left(\exp\left(r\vert \lambda_0\vert_1\right)\right)<\infty,\quad \E\left(\vert Y_0\vert^r_1\right)<\infty.$$  
\end{prop}

\section{Statistical inference}\label{inference}

In this section, we detail our estimation procedures for the dynamic parameters as well as for the copula parameters. 
We first explain the main idea of these methods and introduce our estimators in the two first subsections. Then, we provide some asymptotic results in the two last subsections.

\subsection{Estimation of dynamic parameters}
Going back to the dynamic (\ref{eq::model1}), we now assume that the function $g$ depends on a vector of parameters $\theta\in\Theta\subset\R^Q$ and we denote by $\theta_0$ the vector of parameters associated to a given sample $Y_1,\ldots,Y_n$ generated by (\ref{eq::model1}). 
Under the linear assumption (\ref{eq::linearla}), one can set $\theta=\left(d',\vect(\Gamma)',\vect(A)',\vect(B)'\right)'$ where for a matrix $C$ of any size, $\vect(C)$ denotes the usual vectorization of the matrix $C$.

Inference of parameter $\theta$ can be done by minimizing a suitable contrast.
We construct such contrasts from univariate ones (e.g. log conditional densities). We then adopt a (conditional) pseudo-maximum likelihood approach by writing the likelihood function as if the coordinates of the $Y_t'$s were independent conditionally on their past values. However, the univariate contrasts are not necessarily defined from the conditional log-densities and we provide a more flexible approach by allowing more general univariate contrasts such as least squares or Gaussian quasi-maximum likelihood. 
We recall that $E_1,\ldots,E_k$ denote the state spaces of the different univariate time series, typically one of the sets $\{0,1\}$, $\N$ and $\R$.
Notations $G_1,\ldots,G_k$ are used for the state spaces of the latent processes (typically $G_i=\R$ or $G_i=\R_+$). Finally, for $1\leq i\leq k$, $\mu_i$ will denote either the Lebesgue measure on the real line or the counting measure on $\{0,1\}$ or on $\N$.

To this end, for $1\leq i\leq k$, let $(y,s_i)\mapsto h_{i,y}(s_i)$ be a measurable mapping defined on $E_i\times G_i$ and taking real values such that
\begin{equation}
    \label{eq::classcont}
    \E\left(h_{i,Y}(s_i)\right) \geq \E\left(h_{i,Y}(\overline{s}_i)\right) \text{ and } \E\left(h_{i,Y}(s_i)\right) = \E\left(h_{i,Y}(\overline{s}_i)\right)\Rightarrow s_i=\overline{s}_i
\end{equation} 
whenever $Y\sim P_i\left(\cdot\vert s_i\right)$.
 Here are  two important examples of such functions.
\begin{enumerate}
\item
If the distribution $P_i\left(\cdot\vert s_i\right)$ is absolutely continuous with respect to $\mu_i$, i.e. $P_i\left(dy\vert s_i\right)=p_i\left(y\vert s_i\right)\mu_i(dy)$, one can use the opposite of logarithm of the density $h_{i,y}(\mu)=-\log p_i\left(y\vert \mu\right)$. This case is particularly important when $P_i\left(\cdot\vert s_i\right)$ is specified (e.g. Poisson distribution with parameter $s_i$).
\item
Other standard objective functions such as $h_{i,y}(s_i)=\left(y-s_i\right)^2$ (least-squares estimation) which is adapted to ARMA processes or
$$h_{i,y}(s_i)=\frac{y^2}{\ell(s_i)}+\log\left(\ell(s_i)\right)$$
which corresponds in the context of GARCH type models to Gaussian Quasi-Maximum Likelihood Estimation. Here, the choice $\ell(s_i)=s_i$ corresponds to the standard GARCH model whereas $\ell(s_i)=\exp(s_i)$ corresponds to the log-GARCH model. Note that the Gaussian QMLE can be also used for ARMA type models, setting $h_{i,y}(s_{1,i},s_{2,i})=(y-s_{1,i})^2/s_{2,i}+\log(s_{2,i})$, where the additional parameter $s_{2,i}$ corresponds to the variance of the noise $\varepsilon$, see (\ref{superARMA}).
\end{enumerate}

If $\left(\lambda_t(\theta)\right)_{t\in\Z}$ denotes the process defined recursively by 
$$\lambda_t(\theta)=g_{\theta}\left(\lambda_{t-1}(\theta),Y_{t-1},X_{t-1}\right),\quad \theta\in\Theta,\quad t\in\Z,$$
we define an estimator of $\theta_0$ by minimizing the criterion
\begin{equation}\label{estimautoreg}
\theta\mapsto \ell_n(\theta):=n^{-1}\sum_{t=1}^n\sum_{i=1}^kh_{i,Y_{i,t}}\left(\lambda_{i,t}(\theta)\right),
\end{equation}
where for $t=1,\ldots,n$,
$$\lambda_t(\theta)=g_{\theta}\left(\lambda_{t-1}(\theta),Y_{t-1},X_{t-1}\right).$$
However, as usual with observation-driven model, the previous estimator can not be computed using the available data. To get a feasible estimator, the dynamic of the latent process has to be initialized and we consider a process $\left(\overline{\lambda}_t(\theta)\right)_{t\geq 0}$ defined by $\overline{\lambda}_0(\theta)=\overline{\lambda}_0$ for every $\theta$ in $\Theta$, where $\overline{\lambda}_0$ is a deterministic, and then recursively by $\overline{\lambda}_t(\theta)=g_{\theta}\left(\overline{\lambda}_{t-1}(\theta),Y_{t-1},X_{t-1}\right)$ for $t\geq 1$.
We then define the computable estimator 
\begin{equation}\label{estimautoreg2}
\hat{\theta}=\arg\min_{\theta\in\Theta} n^{-1}\sum_{t=1}^n\sum_{i=1}^kh_{i,Y_{i,t}}\left(\overline{\lambda}_{i,t}(\theta)\right),
\end{equation}
Note that $\hat{\theta}_n$ can be obtained equation by equation for model (\ref{eq::linearla}) with a diagonal matrix $B$. 
Indeed, in this case, parameter $\theta$ is composed of $k$ sub-vectors $\theta^{(1)},\ldots,\theta^{(k)}$ and we have $\hat{\theta}=\left(\hat{\theta}^{(1)},\ldots,\hat{\theta}^{(k)}\right)$ with
$$\hat{\theta}^{(i)}=\arg\min_{\theta^{(i)}} n^{-1}\sum_{t=1}^nh_{i,Y_{i,t}}\left(\overline{\lambda}_{i,t}\left(\theta^{(i)}\right)\right),\quad 1\leq i\leq k.$$
In particular, for $1\leq i\leq k$, $\theta^{(i)}=(d_i,\Gamma(i,1),\ldots,\Gamma(i,m),A(i,1),\ldots,A(i,k),B(i,i))$.

\subsection{Estimation of copula parameters}
Our aim here is to define an estimator of parameter $R_0$ obtained from an estimator of the dynamic parameters.
Note that all the marginal conditional probability distribution $P_i\left(\cdot\vert s_i\right)$, $s_i\in G_i$ are known. 
We assume here that $h_{i,s_i}=-\log p_i\left(\cdot\vert s_i\right)$ for $1\leq i\leq k$. We recall that $p_i(\cdot\vert s_i)$ denotes the probability density of $P_i(\cdot\vert s_i)$ with respect to the measure $\mu_i$.
An estimator of parameter $\theta_0$ can be obtained as explained in the previous section. 
The model being parametric, likelihood inference is adapted for estimating $R_0$. For simplicity, we assume that $F_{1,s_1},\ldots,F_{\ell,s_{\ell}}$ are diffeomorphims (the continuous components)
and $F_{\ell+1,s_{\ell+1}},\ldots,F_{k,s_k}$ are cdf corresponding to discrete distributions with a support included in $\{0,1\}$ or $\N$ (binary or count).
Setting for $1\leq t\leq n$ and $1\leq i\leq k$, $Z_{i,t}(\theta)=F_{i,\overline{\lambda}_{i,t}(\theta)}\left(Y_{i,t}\right)$ and $Z_{i,t}^{-}(\theta)=F_{i,\overline{\lambda}_{i,t}(\theta)}\left(Y_{i,t}^{-}\right)$,
the approximated conditional log-likelihood function for the model is defined by
\begin{eqnarray}\label{vraise}
\ell_n\left(\theta,R\right)&=&\sum_{t=1}^n\log P_R\left(Y_t\vert \overline{\lambda}_t(\theta)\right)\\\nonumber
&=&\sum_{t=1}^n\sum_{i=1}^{\ell}\log p_i\left(Y_{i,t}\vert \overline{\lambda}_{i,t}(\theta)\right)\\\nonumber
&+&\sum_{t=1}^n\log\left\{\int_{Z_{\ell+1,t}^{-}(\theta)}^{Z_{\ell+1,t}(\theta)}\cdots \int_{Z_{k,t}^{-}(\theta)}^{Z_{k,t}(\theta)} c_R\left(Z_{1,t}(\theta),\ldots,Z_{\ell,t}(\theta),u_{\ell+1},\ldots,u_k\right)du_{\ell+1}\cdot du_k\right\}.\nonumber
\end{eqnarray}
Here, $p_R(y\vert s)$ denotes the conditional density of $Y_t$ given $\lambda_t(\theta)=s$ when the copula parameter is $R$ and the autoregressive parameters are given by $\theta$. 

We adopt a plug-in approach by first estimating $\theta_0$ and then optimize the partial log-likelihood function. 
A possible estimator of $R_0$ can be then obtained by minimizing
$$R \mapsto -n^{-1}\ell_n\left(\hat{\theta},R\right),$$
where $\hat{\theta}$ is the estimator obtained as explained in the previous section.

\subsection{Asymptotic results for inference of autoregressive parameters}

In this section, we give a simple set of sufficient conditions ensuring consistency and asymptotic normality of the autoregressive parameters $\theta$. 
\begin{description}
\item[A4]
For $1\leq i\leq k$ and any $s_i^{*}\in G_i$, the mapping
$$\mu\mapsto \int h_{i,y}(s_i)P_i\left(dy\vert s_i^{*}\right)$$
is uniquely minimized at point $s_i=s_i^{*}$.
\item[A5]
For $1\leq i\leq k$, we have 
$$\E\int \sup_{\theta\in \Theta}\left\vert h_{i,y}\left(\lambda_{i,0}(\theta)\right)\right\vert P_i\left(dy\vert \lambda_{i,0}(\theta)\right)<\infty.$$
\item[A6]
For $1\leq i\leq k$, 
$$\frac{1}{n}\sum_{t=1}^n\int \sup_{\theta\in\Theta}\left\vert h_{i,y}\left(\lambda_{i,0}(\theta)\right)-h_{i,y}\left(\overline{\lambda}_{i,0}(\theta)\right)\right\vert P_i\left(dy\vert \lambda_{i,0}(\theta_0)\right)=o_{\P}(1).$$
\item[A7]
We have 
$$\lambda_0(\theta)=\lambda_0(\theta_0)\mbox{ a.s. }\Rightarrow \theta=\theta_0.$$
\end{description}
The proof of the following result is straightforward and follows from standard arguments. See for instance \citep{straumann}, Theorem $5.3.1.$, for the Gaussian QMLE but the arguments used can be extended to this more general setup. 
\begin{theo}\label{consistency}
Let Assumptions {\bf A1-A7} hold true with $\Theta$ a compact subset of $\R^{Q}$. We then have $\lim_{n\rightarrow \infty}\hat{\theta}_n=\theta_0$ a.s.  
\end{theo}
We now turn on the asymptotic normality of our estimator. In what follows, for a function $f:\Theta\rightarrow \R$, we denote by $\nabla f(\theta)$ the gradient vector (column vector of the partial derivatives) and $\nabla^{(2)}f(\theta)$ the Hessian matrix of $f$, evaluated at point $\theta\in\Theta$. 
If $f:\Theta\rightarrow \R^k$, we denote by $J_f(\theta)$ the Jacobian matrix of $f$ at point $\theta\in\Theta$ (we recall that in term of partial derivatives, we have $J_f(\theta)_{i,j}=\frac{\partial f_i}{\partial \theta_j}(\theta)$).
For a function $f$ defined on a subset of the real line, we simply denote by $\dot{f}$ and $\ddot{f}$ its first and second derivatives.
\begin{description}
\item[A8] 
For $1\leq i\leq k$ and $y\in E_i$, the mapping $h_{i,y}$ is two-times continuously differentiable. Moreover, for any $s_i\in G_i$, we have $\int \ddot{h}_{i,y}(s_i)P_i(dy\vert s_i)>0$ and $\int \dot{h}_{i,y}(s_i)P_i(dy\vert s_i)=0$.
\item[A9]
For $1\leq i\leq k$, the random mapping $\theta\mapsto \lambda_{i,0}(\theta)$ is almost surely two-times continuously differentiable and the following uniform integrability condition holds true:
$$\E\int \sup_{\theta\in \Theta}\left[\left\vert \dot{h}_{i,y}\left(\lambda_{i,0}(\theta)\right)\right\vert\cdot\Vert\nabla\lambda_{i,0}(\theta)\Vert^2+\Vert \ddot{h}_{i,y}\left(\lambda_{i,0}(\theta)\right)\nabla^{(2)}\lambda_{i,0}(\theta)\Vert\right]P_i\left(dy\vert \lambda_{i,0}(\theta_0)\right)<\infty.$$
\item[A10]
For $1\leq i\leq k$, we have
$$\E\int \Vert \dot{h}_{i,y}\left(\lambda_{i,0}(\theta_0)\right)\nabla\lambda_{i,0}\left(\theta_0\right)\Vert^2P_i\left(dy\vert\lambda_{i,0}(\theta_0)\right)<\infty.$$
\item[A11]
For $1\leq i\leq k$,
$$\frac{1}{\sqrt{n}}\sum_{t=1}^n \int\sup_{\theta\in\Theta}\Vert \nabla\left(h_{i,y}\circ \lambda_{i,t}\right)(\theta)-\nabla \left(h_{i,y}\circ \overline{\lambda}_{i,t}\right)(\theta)\Vert P_i\left(dy\vert\lambda_{i,t}(\theta_0)\right)=o_{\P}(1).$$
\item[A12]
If there exists $x\in\R^Q$ such that $J_{\lambda_0}(\theta_0)x=0$ a.s. then $x=0$. 
\end{description}
As for consistency, we will not prove the following result. See for instance \citep{straumann}, Theorem $5.6.1.$, the same arguments can be used for proving Theorem \ref{asnormality} below.

\begin{theo}\label{asnormality}
Let Assumptions {\bf A1-A12} hold true with $\theta_0$ being located in the interior of the compact parameter space $\Theta$. We then have
$$\sqrt{n}\left(\hat{\theta}-\theta_0\right)\Rightarrow \mathcal{N}_Q\left(0,J^{-1}IJ^{-1}\right),$$
with 
$$I=\sum_{i=1}^k\sum_{j=1}^k\E\left[\dot{h}_{i,Y_{i,0}}\left(\lambda_{i,0}(\theta_0)\right)\dot{h}_{j,Y_{j,0}}\left(\lambda_{j,0}(\theta_0)\right)\nabla\lambda_{i,0}(\theta_0)\nabla\lambda_{j,0}(\theta_0)'\right],$$
$$J=\sum_{i=1}^k\E\left[\ddot{h}_{i,Y_{i,0}}\left(\lambda_{i,0}(\theta_0)\right)\nabla\lambda_{i,0}(\theta_0)\nabla\lambda_{i,0}(\theta_0)'\right].$$
\end{theo}

Let us note that Assumptions {\bf A8} and {\bf A10} ensure that the process $(M_t)_{t\in\Z}$ defined by $M_t=\sum_{i=1}^k \nabla h_{i,Y_{i,t}}\left(\lambda_{i,t}(\theta_0)\right)$ is a square-integrable martingale difference. Moreover, Assumptions {\bf A8-A9-A12} entail that the Hessian matrix $\mathcal{H}(\theta)=\sum_{i=1}^k\nabla^{(2)}h_{i,Y_{i,0}}\left(\lambda_{i,0}(\theta)\right)$ is well defined, uniformly integrable with respect to $\theta\in\Theta$ and with an invertible expectation at point $\theta_0$. Assumption {\bf A11} guarantees that initializing the latent process has no effect on the asymptotic distribution of the estimator.

\subsection{Sufficient conditions for {\bf A7} and {\bf A12}}\label{identify+}
Here we exhibit a set of simple conditions ensuring both identification of autoregressive parameters and non-degeneracy of the derivative of the latent process. We will provide such conditions for the linear dynamic (\ref{eq::linearla}). Note that the two conditions {\bf A7} and {\bf A12} only involve the autoregressive latent process and not the contrast functions $h_{i,\mu}$. This is why we give a separate study of these two conditions making as few as possible assumptions on the conditional distribution of the multivariate time series model.

In the rest of this section, we assume that 
the trivariate process $\left((Y_t,X_t,U_t)\right)_{t\in\Z}$ is stationary and then that Assumptions {\bf L1-L3} are satisfied.
We recall that $\theta_0=\left(d_0',\vect(\Gamma_0)',\vect(A_0)',\vect(B_0)'\right)'$ denotes the true value of the parameter.
For any $t\in\Z$, we also denote by $\mathcal{F}_t$ the sigma-field generated by $\left(U_j,X_j\right)$, $j\leq t$. 
We will need the following set of assumptions :

\begin{enumerate}
    \item[{\bf I0}] For any $\theta \in  \Theta,~  \rho(B) < 1$, 
    \item[{\bf I1}] For any $v\in \R^m$, we have 
    $$v'X_1\in \mathcal{F}_0\vee \sigma(U_1)\Rightarrow v=0.$$ 
    \item[{\bf I2}] For $1\leq i\leq k$, the function $g_i$ is non-degenerate on the support of $P_i$ and the density $c_{R_0}$ of the copula is positive everywhere.
    \item[{\bf I3}] If $v$ is equal either to a column vector of $A_0$ or to a column vector of $\Gamma_0$, the equalities $B^j v=B_0^j v,\quad j\geq 1$, entail $B=B_0$. 
\end{enumerate}

\begin{lem}
\label{lem::identifiability}
Let Assumptions {\bf I0-I3} hold true for model (\ref{eq::linearla}). Condition {\bf A7} is then satisfied. 
\end{lem}

\paragraph{Notes} 
\begin{enumerate}
\item
For our main setup, the matrices $B$ and $B_0$ are assumed to be diagonal.
In this case, Assumption {\bf I3} is satisfied as soon as all the rows of the concatenated matrix $C:=[A_0,\Gamma_0]$ are non-null. 
\item
Assumption {\bf I1} is more difficult to interpret. It means that any (non degenerate) linear combination of the covariate process at time $t$ cannot 
be explained only by past information and the disturbance term $U_t$. 
For instance, assume that $X_t$ writes as a square integrable infinite moving average expansion $\sum_{j\geq 0}c_j\varepsilon_{t-j}$ where $(c_j)_{j\geq 0}$ is a sequence of matrices and the random vectors $(U_t,\varepsilon_t)$, $t\in\Z$, are i.i.d. In this case, one can take $\mathcal{F}_t=\sigma\left((U_j,\varepsilon_j):j\leq t\right)$ for $t\in\Z$. If $v'X_1$ is measurable with respect to $\mathcal{F}_0\vee \sigma(U_1)$
then so is $v'c_0\varepsilon_1$ which has conditional variance $v'c_0\v\left(\varepsilon_1\vert U_1\right)c_0'v=0$. When $c_0$ is invertible and $\v\left(\varepsilon_1\vert U_1\right)$ is invertible with positive probability, we automatically get $v=0$ and Assumption {\bf I1} is satisfied.  

 Under an additional condition on the covariates, called strict exogeneity, we give below an alternative condition to {\bf I1}.
\end{enumerate}

\begin{enumerate}
\item[{\bf I1'}] The two processes $\left(U_t\right)_{t\in\Z}$ and $\left(X_t\right)_{t\in\Z}$ are independent and if $\sum_{j\geq 1}\Phi_j X_{-j}+c=0$ a.s. then all the matrices $\Phi_j$ of size $p\times m$ and the vector $c$ of length $p$ are equal to zero. 
\end{enumerate}
 The latter condition is satisfied for instance if a linear combination of the coordinates of $X_0$ cannot be equal to an element of $\sigma\left(X_{-j}:j\geq 1\right)$, except if the weights are vanishing. This condition is then the analogue of {\bf I1}, when the noise process and the covariate process are independent. The latter independence condition is often called strict exogeneity in the time series literature.

\begin{lem}\label{lem::identifiability2}
Let Assumptions {\bf I0-I1'-I2-I3} hold true for model (\ref{eq::linearla}). Condition {\bf A7} is then satisfied.  
\end{lem}

The validity of {\bf A12} can be obtained under an additional condition.

\begin{description}
\item[I4] The rank of all the column vectors included in the matrices $B_0^j[A_0,\Gamma_0]$, $j\geq 0$, is equal to $k$. 
\end{description}

\begin{lem}
\label{lem::latentdevfree}
Suppose that either Assumptions {\bf I0-I4} or Assumptions {\bf I0,I1',I2-I4} hold true. Condition {\bf A12} is then satisfied.
\end{lem}

\paragraph{Note.} Assumptions {\bf I3-I4} are checked for instance when the block matrix $[A_0,\Gamma_0]$ is of full rank $k$. However, the latter condition is sufficient but not necessary. For instance if $B$ is diagonal with distinct diagonal elements and the rows of the matrix $C=[A_0,\Gamma_0]$ are all non null, Assumptions {\bf I3-I4} are also satisfied. Indeed in this case, if $v_1,\ldots,v_{k+m}$ denote the column vectors of $C$, the rank of the vectors $B^j v_i$, $0\leq j\leq k$, $1\leq i\leq k+m$, equals to the rank of the matrix $[diag(v_1),\ldots,diag(v_k)]\times I_k\otimes V$, where for $w\in\R^k$, $diag(w)$ denotes the square diagonal matrix with diagonal elements $w_1,\ldots,w_k$, $V$ denotes the (invertible) Vandermonde matrix associated to $B_0(i,i)$, $1\leq i\leq k$, $I_k$ is the diagonal matrix of size $k$ and $\otimes$ denotes the Kronecker product.

\subsection{Examples}
In this section, we go back to our two examples of bivariate time series models.

\subsubsection{Asymptotic results for the GAIN model}
We recall that 
	$\theta = (d', \vect(\Gamma)', \vect(A)', \vect(B)')'$ denotes the vector of parameters we have to estimate
	in the model \eqref{eq::modelGAIN}. When $B$ is assumed to be diagonal, we simply replace $\vect(B)$ by $diag(B)$.
	Here we assume that 
$$\Theta\subset \left\{\theta \in \R_+^Q : \theta_i < 1,\quad Q-1\leq i\leq Q,\quad \min(\theta_1,\theta_2)\geq d_{-}\right\},$$ with 
$Q=2(m+4)$ and $d_{-}$ being a positive constant.
	We combine the Gaussian quasi-likelihood and the Poisson likelihood to estimate the autoregressive parameters, that is $h_{1,y}(\mu_1)=\frac{y^2}{\mu_1}+\log(\mu_1)$ and $h_{2,y}(\mu_2)=\mu_2-y\log(\mu_2)$. 

\begin{prop}
\label{pp::consistencyGAIN}
Consider model \eqref{eq::modelGAIN} with $\Theta\ni \theta_0$ compact. Suppose that Assumption {\bf A1} and Assumptions {\bf I0}, {\bf I1} or {\bf I1'}, {\bf I2-I3} hold true. Suppose furthermore that $\rho\left(A_0+B_0\right)<1$ and that there exists $\delta>0$ such that $\E\vert X_0\vert_1^{1+\delta}<\infty$ and $\E\varepsilon^{2(1+\delta)}<\infty$. The pseudo-maximum likelihood estimator is then strongly consistent, i.e. 
$$\lim_{n\rightarrow \infty}\hat{\theta}_n=\theta_0\mbox{ a.s.}$$
\end{prop}

For asymptotic normality, our result writes as follows.
\begin{prop}
\label{pp::asnormalityGAIN}
Suppose that all the assumptions of Proposition \ref{pp::consistencyGAIN} hold true as well as Assumption {\bf I4}. 
Suppose furthermore that $\theta_0$ belongs to the interior of $\Theta$ and that $\E\varepsilon^4<\infty$.
We then have the convergence in distribution, 
$$\lim_{n\rightarrow \infty}n^{1/2}(\hat{\theta}_n -\theta_0)=\mathcal{N}_Q(0, J^{-1}I {J^{-1}}'),$$
where $I$ and $J$ are given in the statement of Theorem \ref{asnormality}.  
\end{prop}

\subsubsection{Asymptotic results for the BIP model}
 Here, setting $Q=2(m+4)$, we assume that
 $$\Theta\subset\left\{\theta \in \R^Q : \vert \theta_i\vert < 1,\quad Q-1\leq i\leq Q\right\}.$$ 
 We use the pseudo-maximum approach with $h_{1,y}(s_1)=\log\left(1 + e^{s_1}\right) - ys_1$ and $h_{2,y}(s_2)=e^{s_2}-y s_2$.

\begin{prop}
\label{pp::consistencyBIG}
Consider model \eqref{eq::modelBIP} with $\Theta\ni \theta_0$ compact. Suppose that Assumption {\bf A1} and Assumptions {\bf I0}, {\bf I1} or {\bf I1'}, {\bf I2-I3} hold true. Suppose furthermore that $\rho(|B_0|_{vec} + |A_0|_{vec} \mathrm{diag}(1/4, 1)) < 1$,
$||\overline{A}_0|_{vec} + |B_0|_{vec} |_{\infty} < 1$ and that for any $r>0$, $\E\left(\exp\left(r\vert X_0\vert_1\right)\right)<\infty$. The pseudo-maximum likelihood estimator is then strongly consistent, i.e. 
$$\lim_{n\rightarrow \infty}\hat{\theta}_n=\theta_0\mbox{ a.s.}$$
Additionally, if $\theta_0$ is located in the interior of $\Theta$ and if Assumption {\bf I4} holds true, we have asymptotic normality
$$\lim_{n\rightarrow \infty}\sqrt{n}\left(\hat{\theta}_n-\theta_0\right)=\mathcal{N}_Q\left(0,J^{-1}I {J^{-1}}'\right),$$
where $I$ and $J$ are given in the statement of Theorem \ref{asnormality}.  
\end{prop}

\subsection{Asymptotic results for inference of copula parameters}\label{copularesults}

In this subsection, we consider a general parametric model for the copula density. For simplicity, we will only derive consistency results when the initialization of the latent process is ignored, i.e. we identify $\lambda_t(\theta)$ and $\overline{\lambda}_t(\theta)$.
If we assume that the function $g$ in (\ref{eq::model1}) does not depend on its first component (in this case, we use the terminology "pure autoregressive processes"), both processes coincide and our consistency results apply. For non-pure autoregressive processes, deriving a result when the computable version of the latent process is used probably requires more tedious arguments. Note however that such a consistency result seems to be new even in the regression case (i.e. the function $g$ only depends on the exogenous covariates) and it also gives positive results for fitting some existing models to multivariate binary or count times series (\citep{manner}, \citep{DFT}).

As pointed out in \citep{Genest}, it is hopeless to get a systematic identification of the copula parameters when the data are discrete. This is why, we will first state a result showing that one can always estimate consistently the conditional distribution $P_{R_0}\left(\cdot\vert s\right)$ even if identification of the parameter $R$ is not possible. For Gaussian copulas, we next show that such identification is automatic, leading to the consistency of the MLE for the copula parameters.

For $t\in\Z$, we set 
$$f_t\left(\theta,R\right)=\log p_{R}\left(Y_t\vert\lambda_t(\theta)\right),$$
where $p_R\left(\cdot\vert s\right)$, see (\ref{vraise}) for an expression, denotes the density of the conditional distribution $Y_t\vert \lambda_t(\theta)=s$ for a copula parameter $R$ and autoregressive parameters given by $\theta$. 
Note that the conditional distribution of $Y_t$ given $\lambda_t(\theta_0)=s$, denoted by $P_{R_0}\left(\cdot\vert s\right)$, is defined here by
$$P_{R_0}\left(A\vert s\right)=\int_A p_{R_0}(y\vert s)\mu(dy),$$
with $\mu$ being a product of measures with factors equal to either the Lebesgue measure or the counting measure over $\N$ or $\{0,1\}$.
We make the following assumptions.
\begin{description}
\item[A13]
The two parameters $\theta,R$ are contained in some compact sets denoted respectively by $\Theta,\Gamma$.
\item[A14]
The mapping $(\theta,R)\mapsto f_1\left(\theta,R\right)$ is continuous over $\Theta\times \Gamma$ and we have 
$$\E\left(\sup_{(\theta,R)\in\Theta\times\Gamma}\left\vert f_1\left(\theta,R\right)\right\vert\right)<\infty$$
\item[A15]
Setting $\overline{f}(\theta,R)=\E\left(f_1(\theta,R)\right)$, we have
$$\overline{f}\left(\theta_0,R_0\right)\geq \overline{f}\left(\theta_0,R\right),\quad R\in\Gamma.$$
\item[A16]
For any $\lambda$, the mapping $R\mapsto \int\log\left(p_{R}(y\vert \lambda)\right) p_{R_0}(y\vert \lambda)\mu(dy)$ is continuous over $\Gamma$.
\end{description}
Finally let 
$$\mathcal{I}_0=\left\{R\in\Gamma: \overline{f}(\theta_0,R)=\overline{f}(\theta_0,R_0)\right\},$$
$\hat{\theta}$ a strongly consistent estimator of $\theta_0$ and 
$$\hat{R}=\arg\max_{R\in \Gamma}\frac{1}{n}\sum_{t=1}^n f_t\left(\hat{\theta},R\right).$$
In what follows, we denote by $d_{TV}$ the total variation distance, i.e. for two probability measures $\nu$ and $\nu'$ defined on the same measurable space $\left(F,\mathcal{F}\right)$,
$d_{TV}\left(\nu,\nu'\right)=\sup_{A\in \mathcal{F}}\left\vert \nu(A)-\nu(A')\right\vert$. Note that, in the case of existence of a density with respect to the same reference measure $\mu$, i.e. $\nu=f\cdot\mu$ and $\nu'=f'\cdot \mu$, we have the alternative expression
$$d_{TV}\left(\nu,\nu'\right)=\frac{1}{2}\int \left\vert f-f'\right\vert d\mu.$$

\begin{prop}\label{consistCop}
Suppose that Assumptions {\bf A13-A16} hold true with $\hat{\theta}$ a strongly consistent estimator of $\theta_0$. We then have $\lim_{n\rightarrow \infty}d\left(\hat{R},\mathcal{I}_0\right)=0$. Moreover, there exists a Borel set $\Lambda$ such that 
$\P\left(\lambda_0(\theta_0)\in \Lambda\right)=1$ and for any $s\in\Lambda$,
$$d_{TV}\left(P_{\hat{R}}\left(\cdot\vert s\right),P_{R_0}\left(\cdot\vert s\right)\right)\rightarrow 0\mbox{ a.s.}$$
\end{prop}

\paragraph{Note.} Existence of a strongly consist estimator of $\theta_0$ is of course guaranteed from Assumptions {A1-A7}.

We next study the case of Gaussian copula, an important parametric class which is often popular for modeling the joint dependence of continuous or discrete data. See for instance \citep{Marbac}, \citep{Varin} or \citep{DFT}.
In this case, the parameters of the copula can be always identified, even if all the coordinates of the multivariate times series are binary.
However, it is difficult to find in the literature a mathematical study of consistency properties for the estimator of the correlation matrix associated to a Gaussian copula. 
We will provide directly such a result for the multivariate time series models considered in the present paper.
Gaussian copula are defined by 
$$c_R(u_1,\ldots,u_k)=\frac{1}{\sqrt{\det(R)}}\exp\left(-\frac{1}{2}\Phi^{-1}(u)'(R^{-1}-I)\Phi^{-1}(u)\right),$$
where $R$ is a correlation matrix and $\Phi^{-1}(u)=\left(\Phi^{-1}(u_1),\ldots,\Phi^{-1}(u_k)\right)$ with $\Phi^{-1}$ being 
the quantile function of the standard Gaussian distribution.

We will need the following assumptions.

\begin{description}
\item[G1]
The discrete components of the multivariate time series are either binary $\{0,1\}$ or fully supported on $\N$. 
In the latter case, we assume that for any $s_i\in F_i$, $p_i(\cdot\vert s_i)>0$ and 
$$\sum_{y\in\N}\log\left(1-F_{i,s_i}(y)\right)p_i\left(y\vert s_i\right)>-\infty,\quad \sum_{y\in\N}\log\left(p_i(y\vert s_i)\right)p_i(y\vert s_i)>-\infty.$$
\item[G2]
When $1\leq i\leq \ell$ (continuous components), we assume that
$$\E\left[\sup_{\theta\in\Theta}\left\{-\log F_{i,\lambda_{i,0}(\theta)}(Y_{i,0})\right\}+\sup_{\theta\in\Theta}\left\{-\log \left(1-F_{i,\lambda_{i,0}(\theta)}(Y_{i,0})\right)\right\}\right]<\infty.$$
\item[G3]
When $\ell+1\leq i\leq k$ (discrete components), we have to show that
\begin{equation}\label{disc10}
\E\left[\sup_{\theta\in\Theta}\left\{-\log p_i\left(0\vert \lambda_{i,0}(\theta)\right)\right\}+\sup_{\theta\in\Theta}\left\{-\log \left(1-p_i\left(0\vert\lambda_{i,0}(\theta)\right)\right)\right\}\right]<\infty,
\end{equation}
\begin{equation}\label{disc20}
\E\left[\sup_{\theta\in\Theta}\left\{-\log p_i\left(Y_{i,0}\vert \lambda_{i,0}(\theta)\right)\right\}\right]<\infty
\end{equation}
and for count marginal time series,
\begin{equation}\label{disc30}
\E\left[\sup_{\theta\in\Theta}\left\{-\log \left(1-F_{i,\lambda_{i,0}(\theta)}(Y_{i,0})\right)\right\}\right]<\infty.
\end{equation}
\end{description}

\begin{theo}\label{consistGauss}
Suppose that Assumptions {\bf G1-G3} hold true with $\Theta\times\Gamma$ compact and that $\hat{\theta}$ is a strongly consistent estimator of $\theta_0$.
We then have strong consistency of the two-step estimator, i.e. $\lim_{n\rightarrow \infty}\hat{R}=R_0$ a.s.
\end{theo}

\paragraph{Note.} The additional assumptions {\bf G1-G3} are not so restrictive. For instance, Poisson or logistic autoregressive models and GARCH or ARMA models will satisfy these conditions in general (up to some additional regularity conditions on the noise density).
Below, we carefully check these assumptions for the GAIN and BIN model.

\subsubsection{Consistency for the GAIN model}

Here, the correlation matrix writes as $R_0=\begin{pmatrix} 1&r_0\\r_0&1\end{pmatrix}$ and $r_0$ is the single parameter to estimate. 
For simplicity, we give below a consistency result when the noise of the GARCH component is Gaussian, though other probability distributions are possible.

\begin{cor}\label{contentGAIN}
Suppose that all the assumptions of Proposition \ref{pp::consistencyGAIN} are valid and that the noise $\epsilon$ for the GARCH component is $\mathcal{N}(0,1)-$distributed.
We then have 
$$\lim_{n\rightarrow \infty}\hat{r}=r_0\quad{ a.s.}$$
\end{cor}

\subsubsection{Consistency for the BIP model}

Consistency also holds for the BIP models when the assumptions ensuring consistency of the pseudo-maximum likelihood estimator are satisfied.

\begin{cor}\label{contentBIP}
Suppose that all the assumptions of Proposition \ref{pp::consistencyBIG} are satisfied.
We then have 
$$\lim_{n\rightarrow \infty}\hat{r}=r_0\quad{ a.s.}$$
\end{cor}

\section{Numerical experiments and real data applications}\label{numeric}

In this section, we discuss the implementation of our inference procedure for the GAIN and the BIP model. We only implement these models for a Gaussian copula.
There also exist many other interesting families of copula (Clayton, Gumbell...) and we defer the reader to \citep{bouye} for an interesting survey about copulas properties and their use in finance. Throughout this section, the density of the GARCH noise is always assumed to be a standard Gaussian.

The main difficulty for fitting our models is the approximation of the likelihood function for estimating the correlation matrix $R_0$. Pseudo-likelihood estimation of autoregressive parameters is straightforward. Note that the equation-by-equation estimation can be obtained from the standard software packages since it is equivalent to fit a standard time series model to one coordinate with past values of the other coordinates as covariates.    
When $k=2$, we get a simpler formula for the likelihood function (\ref{vraise}). In particular, the correlation matrix $R_0$ only involves one coefficient $r_0\in (-1,1)$ and using the properties of conditional distributions for Gaussian vectors, one can show that an estimation of parameter $r_0$ can be obtained by minimizing
$$r\mapsto \sum_{t = 1}^{n} \log\left( \left\{ \Phi\left(\frac{\Phi^{-1}(Z_{i,t}) - r \Phi^{-1}(Z_{j,t})}{\sqrt{1-r^2}}\right) - \Phi\left(\frac{\Phi^{-1}(Z_{i,t}^-) - r \Phi^{-1}(Z_{j,t})}{\sqrt{1-r^2}}\right)\right\}\right)  $$ 
for the GAIN model and 
$$r\mapsto \sum_{t = 1}^{n} \log\left(\int_{0}^{1} \left\{ \Phi\left(\frac{\Phi^{-1}(Z_{i,t}) - r \epsilon_{j,t}(u)}{\sqrt{1-r^2}}\right) - \Phi\left(\frac{\Phi^{-1}(Z_{i,t}^-) - r \epsilon_{j,t}(u)}{\sqrt{1-r^2}}\right)\right\}du\right)
$$ 
for the BIP model. Here, $Z_{i,t} = F_{i, \hat{\lambda}_{i,t}}(Y_{i,t})$, $Z_{i,t}^- = F_{i, \hat{\lambda}_{i,t}}(Y_{i,t}-1)$ and $\epsilon_{i,t}(u_i) = \Phi^{-1}\left(Z_{i,t} - u_i(Z_{i,t} - Z_{i,t}^-)\right)$, where $\hat{\lambda}_t=\overline{\lambda}_t\left(\hat{\theta}\right)$.

Note that for the GAIN model, the formula is explicit in term of the Gaussian cdf $\Phi$ whereas the formula for the BIP model involves the computation of one integral. Approximation for this integral can be obtained from Monte Carlo methods. In our simulations, we simply simulate a sample of size $N=10^4$ of uniformly distributed random variables and approximate this integral by an empirical counterpart. Note that when $k\geq 3$, several iterated integrals have to be computed for approximating (\ref{vraise}) and it is hopeless to get an accurate approximation of the likelihood function using the previous method. To overcome this problem, one can use the importance sampling strategy considered in \citep{Varin}. When $k=2$, there is no gain in applying this method. When $k\geq 3$, there is also the possibility to use pairwise composite likelihood methods as discussed in \citep{Varin}. In this paper, we will not investigate such computational issues and their corresponding convergence properties. 

One can also compute standard errors for our parameters. For the autoregressive parameters, the asymptotic distribution of pseudo-likelihood estimators can be used. For the copula parameter, we did not investigate the asymptotic distribution of the likelihood estimator. However, one can simply use a parametric bootstrap: we simulate $B$ paths of size $n$ of the model using the estimated parameters $\hat{\theta}$ and $\hat{r}$, we then compute the standard error from the sample  $\hat{r}^{*,b}$ for $b=1,\ldots,B$. A theoretical justification of such a procedure is beyond the scope of this paper.

\subsection{Numerical experiments}
We fitted the GAIN and BIP models to simulated data. For the BIP model, we used an additional covariate process with $m=1$ and defined by an AR$(1)$ process, 
$X_t=-0.15\times X_{t-1}+\xi_t$ where $\left(\xi_t\right)_{t\in\Z}$ is a sequence of i.i.d. standard Gaussian random variables. This sequence is assumed to be independent of the sequence $\left(U_t\right)_{t\in\Z}$ used for the copula. For both models, Tables \ref{1},\ref{2},\ref{3} and \ref{4} give averages and standard deviations of $M=500$ estimators and for two sample sizes, $n=500$ and $n=1000$. One can note that both estimation of autoregressive parameters and of the copula parameter work reasonably well whatever the values of $r_0$ which is allowed to vary from $-0.9$ to $0.9$. We found that $n=500$ is a reasonable sample size to get an accurate estimation of all the parameters.

\subsection{An application to sleep data}
We use the data set already studied in \citep{fokianos2003}, with sleep state measurements of a newborn infant together with his heart rate $Y_{2,t}$ (taking integer values) and temperature $X_t$ sampled every $30$ seconds. The sample size is $n=1024$ and the sleep states are classified as: $(1)$ quiet sleep, $(2)$ indeterminate sleep, $(3)$ active sleep, $(4)$ awake. To define a binary time series, we aggregate States $(1)$, $(2)$ and $(3)$ and we then set $Y_{1,t}=1$ when the infant is awake and $0$ if it is not. A BIP model is fitted to the time series $(Y_t)_{1\leq t\leq n}$. It is quite intuitive to suspect a dependence between the heart rate and the sleep state and our aim is to analyze such a joint dynamic.

Results are displayed in Table \ref{sleepdata}. We consider both a fitting of the full BIP model $(I)$ and of a restricted model $(II)$ without the temperature as an exogenous covariate. Using $t-$tests, one can note that the lag value of the temperature seems not to have a significant contribution to the dynamic, though the AIC is smaller when this covariate is incorporated in the model. Both lag values of the heart rate (the sleep state respectively) seem to have a negligible influence to the present value of the sleep state (the heart rate respectively).
On the other hand, we get a positive coefficient $\hat{r}$ for the copula and we then observe a positive association between the two time series at time $t$.
Being awake is more likely associated with larger heart rates at the same time which is a quite logical.
Note that such findings are compatible with that of the univariate modeling of \citep{fokianos2003} (see Table $10$) with a sleep state at time $t$ which seems to depend on the current heart rate but less on its lag value. 

\begin{table}[]
	\small
	\centering
	\begin{tabular}{|ccccc|ccccc|}
		\hline
		\multicolumn{5}{|c|}{Log-Poisson (I)} & \multicolumn{5}{|c|}{Log-Poisson (II)}  \\ \hline
		$d_1$ & $A(1,1)$ & $A(1,2)$ & $B(1,1)$ & $\Gamma(1,1)$  & $d_1$ & $A(1,1)$ & $A(1,2)$ & $B(1,1)$ & $\Gamma(1,1)$  \\ \hline
		-10.8984   &  0.6929 &  0.0068  &  0.0289 &   3.3907  & 1.0859 &  0.7437 &  0.0108 &  0.0331  &  \\ 
		(33.4872) &  (0.0428) &  (0.8275) & (0.0907) & (9.3608) & (0.1611) & (0.1380) &  (1.1646) & (0.1328) &  \\ \hline
		\multicolumn{5}{|c|}{Logit-Binary (I)} & \multicolumn{5}{|c|}{Logit-Binary (II)} \\ \hline
		$d_2$ & $A(2,1)$ & $A(2,2)$ & $B(2,2)$ & $\Gamma(2,1)$ & $d_2$ & $A(2,1)$ & $A(2,2)$ & $B(2,2)$ & $\Gamma(2,1)$ \\ \hline
		582.3259  &   -1.5590 &   13.9541  &    -0.5054  &  -161.0862 &  7.1203 & -2.9145 &  13.0971 &  -0.4675  &   \\ 
		(503.0512)  & (3.0956)  & (1.4844)  & (0.1090) & (138.3401)  & (16.4880) &  (3.3355) &   (1.3732)   & (0.1171) &  \\ \hline
		\multicolumn{5}{|c|}{$r$ (I)} & \multicolumn{5}{|c|}{$r$ (II)} \\ \hline
		estimate : & 0.3337  &  sd :  & 0.1040  &   & estimate : & 0.2749 &    sd : & 0.1058     & \\ \hline 
		\multicolumn{5}{|l|}{ (I) :  AIC = -535318.3} & \multicolumn{5}{|l|}{ (II) :  AIC = -535309.7 } \\ \hline
		\end{tabular}
		\caption{Estimation of the parameters of the BIP model for sleep data. Standard errors are given in parenthesis.}
	\label{sleepdata}
\end{table}

\subsection{An application to high-frequency transactions in finance}
The data are downloaded from \url{http://www.nasdaqomxnordic.com} and represent the real-time transactions on Boliden, a metal exploring, extracting and processing firm. The count component, denoted by $Y_{2,t}$, is the number of transactions of this stock occurring in a time interval of two successive minutes. The continuous coordinate is the log-return $Y_{1,t}=\log\left(P_t\right)-\log\left(P_{t-1}\right)$ of the transaction average price $P_t$. The transaction average price $P_t$ is simply given by $\sum_{j=1}^m W_{j,t}P_{j,t}/\sum_{j=1}^m W_{j,t}$ where $W_{j,t}$ is the number of transactions at price $P_{j,t}$ occurring during this two minutes time interval. See also \citep{FokTjo} who used a similar weighted average price. 
We model the dynamic of $\left(Y_t\right)_{1\leq t\leq 467}$ with a GAIN process.  The data are collected for a two days time period between March $29$th and March $30$th , $2021$. The result are given Table \ref{Transactions}. For the autoregressive parameters, one can suspect that the lag value of the number of transactions has no effect on the volatility and then on the next log-return. We then test the hypothesis $H_0$: $A(1,2)=0$ versus $H_1$: $A(1,2)>0$. Since the parameter is on the boundary of the parameter set under the null hypothesis, the QMLE has not an asymptotic Gaussian distribution. We then used the corrected test given in \citep{Francq} which consists in rejecting $H_0$ at level $\alpha$ if $\hat{A}(1,2)^2/\hat{v}$ is larger than the quantile of order $1-2\alpha$ (instead of $1-\alpha$ when the parameter is not on the boundary) of a $\chi^2$ distribution with $1$ degree of freedom. Here $\hat{v}$ is simply an estimation of the asymptotic variance of $\hat{A}(1,2)$ given in Proposition \ref{pp::asnormalityGAIN}.
We do not reject $H_0$ at level $\alpha=5\%$. Moreover, $\hat{r}$ is quite small and negative. Unfortunately, we did not derive the asymptotic distribution of this estimator to get a standard significance test. If the asymptotic distribution was Gaussian at the usual $\sqrt{n}$ convergence rate, we would reject the hypothesis $H_0$: $r=0$, but further investigation is needed to make a rigorous conclusion.
One can then conclude that past values of the log-returns seem to have an influence on the number of transactions at time $t$ but not the inverse. Moreover a negative but very small association between the two time series at time $t$ is possible.    

\begin{table}[]
	\small
	\centering
	\begin{tabular}{|cccc|}
		\hline
		\multicolumn{4}{|c|}{GARCH}   \\ \hline
		$d_1$ & $A(1,1)$ & $A(1,2)$ & $B(1,1)$   \\ \hline
		0.0012 &  0.7569 &  0.00005 & 0.1226   \\ 
		(0.0353) &  (0.3829) &  (0.0005) &  (0.0017)   \\
		\hline
		\multicolumn{4}{|c|}{INGARCH} \\ \hline
		$d_2$ & $A(2,1)$ & $A(2,2)$ & $B(2,2)$      \\ \hline
		3.1988  & 34.6621 &   0.1236 &   0.7719   \\ 
		(1.2767) &  (7.3539) &  (0.0406) &  (0.0702) \\ \hline
		\multicolumn{4}{|c|}{$r$   }  \\ \hline
		estimate :   & -0.011 & sd : & 0.0049   \\ \hline
		\multicolumn{4}{|l|}{ (I) :  AIC = -40449.32} \\ \hline
	\end{tabular}
	\caption{Estimation of the parameters of the GAIN model for financial data. Standard errors are given in parenthesis.}
	\label{Transactions}
\end{table}

\section{Proofs of the results}\label{proof}

\subsection{Proof of the results of Section \ref{autoreg}}

\subsubsection{Proof of Theorem \ref{th::stabilityLatent}}
Define the mapping
$$f_t:s\mapsto g\left(s,F_s^{-1}(U_{t-1}),X_{t-1}\right).$$
From {\bf A1-A3}, the assumptions of Theorem $4$ in \citet{DT} are satisfied with $o=p=1$ and $\zeta_t=\left(X_{t-1},U_{t-1}\right)$.
In particular, there exists a unique stationary, integrable and $\left(\mathcal{F}_{t-1}\right)_{t\in\Z}-$ adapted process $\left(\lambda_t\right)_{t\in\Z}$
such that $\lambda_t=f_t\left(\lambda_{t-1}\right)$. Moreover, Theorem $2$ in \citet{DT} guarantees the representation $\lambda_t=H\left(\left(U_{t-j},X_{t-j}\right)_{j\geq 1}\right)$ for a suitable measurable function $H$ defined on an infinite Cartesian product (Bernoulli shift representation with respect to the process $((X_t,U_t))_{t\in\Z}$). Setting $Y_t=F_{\lambda_t}^{-1}(U_t)$, we then get a stationary and ergodic process $\left((Y_t,\lambda_t,X_t)\right)_{t\in\Z}$, as this process also has a Bernoulli representation with respect to the stationary and ergodic process $\left((X_t,U_t)\right)_{t\in\Z}$.
The uniqueness property easily follows.$\square$

\subsubsection{Proof of Proposition \ref{pp::stationaryGAIN}}
\begin{enumerate}
\item
The first point is a consequence of Corollary \ref{cor::stabilityLinear} with $g_1(y)=y^2$ and $g_2(y)=y$. 
Assumption {\bf L1} is easy to check and from the discussion given in Section \ref{discussion}, Assumption {\bf L2} is also satisfied for both coordinates with $c_1=c_2=1$.
Straightforwardly, $\E(|\overline{Y}_0|_1)  < \infty$ since $\E(|\overline{Y}_0|_1) = \E(|\lambda_0|_1)$.
\item
Under the proposed assumptions, we have 
$$\E\lambda_t=d+B\E\lambda_{t-1}+A\E \overline{Y}_{t-1}+\Gamma \E X_{t-1}=d+\Gamma\E X_{t-1}+(A+B)\E\lambda_{t-1}.$$
By stationarity, we get $m:=\E\lambda_0=c+(A+B)m$ where $c=d+\Gamma \E X_0$. Iterating the previous equality, we get $m=\sum_{i=0}^K(A+B)^ic$. By positivity of the components of $c$ and non negativity of the matrices $A$ and $B$, we deduce that the series $\sum_{i=0}^K(A+B)^i$ is converging term by term and then that $\lim_{i\rightarrow \infty}(A+B)^i=0$. This automatically imply that $\rho(A+B)<1$.$\square$ 

\end{enumerate}
\subsubsection{Proof of Proposition \ref{pp::momentGAIN}}

For a positive real $s$ and an integer $r \geq 1,$
\begin{enumerate}
    \item $\E^{1/r}\left[F^{-1}_{1,s}(U_{1,t})^{2r}\right] = \E^{1/r}\left[\epsilon^{2r}\right] s $
    \item $\E^{1/r}\left[F^{-1}_{2,s}(U_{2,t})^{r}\right] \leq  (1+\delta) s + b_{r, \delta}$
\end{enumerate}
for $\delta > 0$ arbitrarily small and  $ b_{r, \delta}$ a positive constant which depends on $r$ and $\delta$ (see Lemma \ref{lem::poism} below).
We will apply Lemma \ref{lem::momFonc}. Since
\begin{eqnarray*}
\|g(s, F_s^{-1}(U_t), X_t)\|_{r, t-1, vec} & \preccurlyeq & (A\mathcal{H} + B )s + \Gamma \Vert X_t\Vert_{r,t-1,vec} + \overline{b}_{r, \delta} 
\end{eqnarray*}
where $\overline{b}_{r, \delta}   = (0, {b}_{r, \delta}  )$ and $\mathcal{H} = \mathrm{diag}(\E^{1/r}[\epsilon ^ {2r}], 1 + \delta)$, the result follows from Lemma \ref{lem::momFonc}. Indeed, if $\delta$ is small enough, we have $\rho\left(A\mathcal{H}+B\right)<1$.
$\square$

\subsubsection{Proof of Proposition \ref{pp::stationaryBIG}} 
Setting $g_1(y)=y$ and $g_2(y)=\log(1+y)$, the discussion given in Section \ref{discussion} shows that {\bf L2} is satisfied with $c_1=1/4$ and $c_2=1$.
Assumption {\bf L1} is straightforward to show. The result then follows from an application of Corollary \ref{cor::stabilityLinear}. The integrability condition
follows from the fact that $(Y_{1,t})_{t\in\Z}$ is bounded and the inequality $\E\log(1+Y_{2,t})\leq \E\vert \lambda_{2,t}\vert<\infty$ which follows from the discussion of Section \ref{discussion}.$\square$

\subsubsection{Proof of Proposition \ref{pp::momentBIG}}
We will apply the result of lemma \ref{lem::momFonc} with the $r=1$ and the function $\phi(s)=\exp(\kappa \vert s\vert_{vec})$ where the exponential function is applied componentwise. Setting $\mu_i=\sum_{j=1}^2\left(\vert \overline{A}(i,j)\vert+\vert B(i,j)\vert\right)<1$, $C_i=\vert A(i,1)\vert$ for $i=1,2$ and denoting by $\odot$ the Hadamard product, we have 
$$\vert g(s,F_s^{-1}(U_t),X_t)\vert_{vec}\preccurlyeq (1-\mu)\odot \frac{\vert d+\Gamma X_t+C\vert_{vec}}{1-\mu}+\vert \overline{A}\vert_{vec}\begin{pmatrix} 0\\F_{2,s_2}^{-1}(U_{2,t})\end{pmatrix}+\vert B\vert_{vec}\vert s\vert_{vec}.$$
Using convexity of the exponential function, we deduce that
\begin{equation}\label{expmom}
\phi\left(g(s,F_s^{-1}(U_t),X_t)\right)\preccurlyeq (1-\mu)\odot \phi\left(\frac{d+\Gamma X_t+C}{1-\mu}\right)+\vert \overline{A}\vert_{vec}\begin{pmatrix} 1\\\exp(\kappa g_2\circ F_{2,s_2}^{-1}(U_{2,t}))\end{pmatrix}+\vert B\vert_{vec}\phi(s).
\end{equation}
In what follows, we denote by $c_{t-1}$ the conditional expectation of $(1-\mu)\odot\phi\left(\frac{d+\Gamma X_t+C}{1-\mu}\right)$ with respect to $\mathcal{F}_{t-1}$. Note that for any $\delta>0$, there exists $d_{\kappa,\delta}>0$ such that
$$\E\exp(\kappa g_2\circ F_{2,s_2}^{-1}(U_{2,t}))=\E\left(1+F_{2,s_2}^{-1}(U_{2,t})\right)^{\kappa}\leq (1+\delta)\exp(\kappa s_2)+d_{\kappa,\delta}.$$
The previous bound can be obtained from Lemma \ref{lem::poism}, using the convexity of power functions. 
Taking the conditional expectation with respect to $\mathcal{F}_{t-1}$ in (\ref{expmom}), we deduce that 
$$\Vert \phi\left(g(s,F_s^{-1}(U_t),X_t)\right)\Vert_{t-1,1,vec}\preccurlyeq c_{t-1}+\left(\vert \overline{A}\vert_{vec}diag(1,1+\delta)+\vert B\vert_{vec}\right)\phi(s).$$
Taking $\delta$ small enough, our assumptions guarantee that the spectral radius of $\vert \overline{A}\vert_{vec}diag(1,1+\delta)+\vert B\vert_{vec}$ is less than $1$ and Lemma \ref{lem::momFonc} leads to the results.$\square$

\subsubsection{Proof of Lemma \ref{lem::identifiability}}

Suppose that $\lambda_t(\theta)=\lambda_t(\theta_0)$ a.s. 
Since $\forall B \in \Theta, \rho(B) < 1$, we get 
\begin{eqnarray}
\label{eq::operatoridenteq}
    \sum_{j=1}^{\infty}\left[B^{j-1}A-B_0^{j-1}A_0\right]\overline{Y}_{t-j}& =  & \sum_{j=1}^{\infty}\left[B^{j-1}\Gamma-B_0^{j-1}\Gamma_0\right]X_{t-j}  \nonumber\\
     & & + \sum_{j=1}^{\infty}\left[B^{j-1}d-B_0^{j-1}d_0\right].
\end{eqnarray}
and consequently there exist a set of matrices $\Psi_j, \Phi_j, j\geq 1$ and a vector $c$ of $\R^p$ such that 
\begin{equation}\label{goodinter}
\sum_{j \geq 1} \Psi_j \overline{Y}_{t-j} = c +  \sum_{j \geq 1} \Phi_j X_{t-j}\mbox{ a.s.}
\end{equation}
From ${\bf I1}, \Phi_1 = 0$. Indeed, under our assumptions, if the previous equality is valid, the random vector $\Phi_1 X_{t-1}$ is measurable with respect to the sigma-field $\mathcal{F}_{t-2}\vee \sigma(U_{t-1})$. 

Next, suppose that $\Psi_1\neq 0$. There then exists a vector $v\in\R^p\setminus\{0\}$ such that $v'\overline{Y}_{t-1}=G_{t-2}$, where $G_{t-2}$ is a random variable $\mathcal{F}_{t-2}-$measurable.
For $1\leq i\leq p$, set $H_i=g_i\circ F_{i,\lambda_{i,t-1}(\theta_0)}^{-1}$. 
Note that
$$1=\P\left(v'\overline{Y}_{t-1}=G_{t-2}\vert \mathcal{F}_{t-2}\right)=\int_{[0,1]^p}c_{R_0}(u_1,\ldots,u_p)\mathds{1}_{\sum_{i=1}^p v_iH_i(u_i)=G_{t-2}}du_1\cdots du_p.$$
Since $c_{R_0}$ is positive, we deduce that $\lambda_p\left(\left\{\sum_{i=1}^p v_i H_i(\cdot)=G_{t-2}\right\}\right)=1$ a.s. where $\lambda_p$ denotes the Lebesgue measure 
on $[0,1]^p$. From ${\bf I2}$, we automatically have $v=0$ because otherwise, the value of one of the $H_i'$s is determined by the values of the others functions $H_j$. We then get $\Psi_1 = 0$. 
Recursively, we obtain $\Phi_j = \Psi_j = 0, \forall j \geq 2$ and finally $c = 0$. 

Then, the equation \eqref{eq::operatoridenteq} yields $B^j A=B_0^j A_0$ for any $j\in\N$ and then $A=A_0$. Moreover, $B^j\Gamma=B_0^j\Gamma_0$ for all $j\in\N$ entails $\Gamma=\Gamma_0$. 
From {\bf I3}, we get $B=B_0$ and then $d=d_0$. $\square$

\subsubsection{Proof of Lemma \ref{lem::identifiability2}}

The proof is similar to the proof of Lemma \ref{lem::identifiability}. The single difference concerns the treatment of the equality (\ref{goodinter}).
If the noise process and the covariate process are independent, then the conditional distribution of $Y_t$ given $\mathcal{F}_{t-1}$ is also the conditional distribution 
of $Y_t$ given $\sigma\left(X_j: j\in\Z\right)\vee \sigma\left(U_{t-i}:i\geq 1\right)$. Assume an equality of the form (\ref{goodinter}). From {\bf I2}, we obtain recursively $\Psi_j=0$ for $j\geq 1$, using the same arguments as in the proof of Lemma \ref{lem::identifiability}.
Hence, we get $\sum_{j\geq 1}\Phi_j X_{t-j}+c=0$ a.s. From {\bf I1'}, we get $\Phi_j=0$ for $j\geq 1$ and then $c=0$. The rest of the proof is identical to that of Lemma \ref{lem::identifiability}.$\square$ 

\subsubsection{Proof of Lemma \ref{lem::latentdevfree}}

For $1\leq i,j\leq k$, $1\leq i'\leq k$ and $\leq j'\leq m$, let us denote by $E_{i,j}$ and $G_{i',j'}$ the matrices of size $k\times k$ and $k\times m$ respectively and with elements equal to $1$ for the couple of indices $(i,j)$ or $(i',j')$ and $0$ elsewhere.
We also denote by $J_{\lambda_t}(\theta_0)$ the Jacobian matrix of $\lambda_t$ at point $\theta_0$. 
Assume that there exists a vector $x$ such that $J_{\lambda_t}(\theta_0)x=0$ a.s.
We have 
$$J_{\lambda_t}(\theta_0)=[1|\overline{Y}_{t-1}|X_{t-1}|\lambda_{t-1}(\theta_0)]+B_0 J_{\lambda_{t-1}}(\theta_0),$$
where $[1|\overline{Y}_{t-1}|X_{t-1}|\lambda_{t-1}(\theta_0)]$ is a concatenated matrix with block elements
$$I_k,E(1,1)\overline{Y}_{t-1},\ldots, E(k,k)\overline{Y}_{t-1},G(1,1)X_{t-1},\ldots,G(k,m)X_{t-1},E(1,1)\lambda_{t-1}(\theta_0),\ldots, E(k,k)\lambda_{t-1}(\theta_0),$$
with $I_k$ the identity matrix of size $k$. 
By stationarity, we also have $J_{\lambda_{t-1}}(\theta_0)x=0$ a.s. and we then obtain $[1|\overline{Y}_{t-1}|X_{t-1}|\lambda_{t-1}(\theta_0)]x=0$. 
We then deduce the existence of a vector $\nu$ in $\R^k$, two square matrices $\alpha$ and $\beta$ of size $k$ and a matrix $\gamma$ of size $k\times m$
such that 
$$\nu+\alpha \overline{Y}_{t-1}+\beta \lambda_{t-1}(\theta_0)+\gamma X_{t-1}=0\mbox{ a.s.}$$
Using the same kind of arguments as in the proof of Lemma \ref{lem::identifiability} (see the implication of equality (\ref{goodinter})), we get the equalities $\nu+\beta(I-B_0)^{-1}d_0=0$, $\alpha=0$, $\gamma=0$ and $\beta B_0^j[A_0,\Gamma_0]=0$ for any $j\geq 0$. From {\bf I4}, we get $\beta=0$ and then $\nu=0$. 
Since $x=\left(\nu',\vect(\alpha)',\vect(\gamma)',\vect(\beta)'\right)'$, we get $x=0$ which means that {\bf A12} is satisfied.$\square$

\subsection{Proofs of the results of Section \ref{inference}}

\subsubsection{Proof of proposition \ref{pp::consistencyBIG}}
We check the assumptions of Theorem \ref{consistency} and Theorem \ref{asnormality}.
Let us denote $h_{1,Y_{1,t}}(\lambda_t(\theta)) = \log(1 + e^{\lambda_{1,t}(\theta)}) - Y_{1,t}\lambda_{1,t}(\theta)$, $h_{2,Y_{2,t}}(\lambda_t(\theta)) = e^{\lambda_{2,t}(\theta)} - Y_{2,t}\lambda_{2,t}(\theta)$.
{\bf A4} and {\bf A8} are straightforward to show or result from the Kullback-Leibler divergence properties. Moreover {\bf A7} and {\bf A12} follows from the results given in Subsection \ref{identify+}.
We next check {\bf A5}. We have
$$\E\left(\sup_{\theta\in \Theta}\left\vert h_{1,Y_{1,t}}\left(\lambda_{1,t}(\theta)\right)\right\vert\right)\leq \log(2)+2\E\left(\sup_{\theta\in\Theta}\vert \lambda_{1,t}(\theta)\vert\right)$$
and 
$$\E\left(\sup_{\theta\in \Theta}\left\vert h_{2,Y_{2,t}}\left(\lambda_{1,t}(\theta)\right)\right\vert\right)\leq \exp\left(\sup_{\theta\in\Theta}\vert \lambda_{2,t}(\theta)\vert\right)+\exp\left(\lambda_{2,t}(\theta_0)\right)\sup_{\theta\in\Theta}\vert \lambda_{2,t}(\theta)\vert.$$
By recursion, note that
$$
\lambda_t(\theta) = (I-B)^{-1} ( d + \Gamma X_{t-1}) + \sum_{j \geq 0} B^j A \overline{Y}_{t-j-1}
$$
and then 
$$
 \sup_\theta |\lambda_t(\theta)|_1 \leq \sup_\theta |(I-B)^{-1} d|_1  + \sup_\theta |(I-B)^{-1}\Gamma X_{t-1}|_1 +  \sup_\theta |A|_1  \sum_{j\geq 0} |B^j|_1|\overline{Y}_{t-j-1}|_1.
$$
From lemma \ref{lem::bornerayon}, there exits $\tau \in (0,1)$ and $\kappa > 0$ such that :
\begin{equation}
    \label{eq::expsuplambda}
    \E(e^{\sup_\theta |\lambda_t(\theta)|_1 }) \leq K (1-\tau)\sum_{j\geq 0} \tau^j\E\left(e^{\frac{ \kappa \sup_\theta |A|_1}{1-\tau}  |\overline{Y}_{t-j-1}|_1}\right) = K\E\left(e^{\frac{ \kappa \sup_\theta |A|_1}{1-\tau}  |\overline{Y}_{0}|_1}\right) 
\end{equation}

with $K = e^{ \sup_\theta |(I-B)^{-1} d|_1} \E\left(e^{ \sup_\theta |(I-B)^{-1}\Gamma X_{0}|_1}\right)$. From Proposition \ref{pp::momentBIG}, 
\begin{equation}
    \label{eq::expy}
    \E\left(e^{\frac{ \kappa \sup_\theta |A|_1}{1-\tau}  |\overline{Y}_{0}|_1}\right) \leq \left(1+e^{\frac{ \kappa \sup_\theta |A|_1}{1-\tau}}\right) \E\left[\left(1 + {Y}_{2,0} \right)^{\frac{\kappa \sup_\theta |A|_1}{1-\tau}}\right] < \infty.
\end{equation}
 Altogether, $\sup_{\theta\in\Theta} |\lambda_t(\theta)|_1$ admits a finite exponential moment of any order and $Y_{2,0}$ has all polynomial moments. On other hand, $\sup_\theta e^{\lambda_{2,t}(\theta)} \leq  e^{\sup_\theta \lambda_{2,t}(\theta)}$ since exponential function is increasing. Then {\bf A5} follows.

We next check {\bf A6}. 
Using the Lipschitz property of the function $h_{1,t}, h_{2,t}$, we only have to show that 
$$\sum_{t\geq 1}\left(\sup_{\theta\in \Theta}\vert \lambda_{1,t}(\theta)-\overline{\lambda}_{1,t}(\theta)\vert+\exp\left(\sup_{\theta\in\Theta}\vert \lambda_{2,t}(\theta)\vert\right)\sup_{\theta\in\Theta}\vert \lambda_{2,t}(\theta)-\overline{\lambda}_{2,t}(\theta)\vert\right)<\infty.$$
Then {\bf A6} follows from the existence of exponential moments and Lemma \ref{lem::approxlatent}.

Next, we check {\bf A9-A10}. To this end, it is sufficient to show that the random variables
$$\sup_{\theta\in\Theta}\Vert \nabla \lambda_{1,0}(\theta)\Vert^2,\quad \sup_{\theta\in\Theta}\Vert \nabla^{(2)}\lambda_{1,0}(\theta)\Vert,\quad \exp\left(2\sup_{\theta\in\Theta}\vert \lambda_{2,0}(\theta)\vert\right)\times \sup_{\theta\in\Theta}\vert \nabla\lambda_{2,0}(\theta)\vert$$
and
$$\exp\left(\sup_{\theta\in\Theta}\vert\lambda_{2,0}(\theta)\vert\right)\times \sup_{\theta\in \Theta}\Vert \nabla^{(2)}\lambda_{2,0}(\theta)\Vert.$$
These integrability conditions follows from the existence of exponential moments and Lemma \ref{lem::momlatent}.

Finally, we check {\bf A11}. This condition will be satisfied as soon as $\sup_{t\geq 1}d_t<\infty$ with $d_t$ being equal to one of the following quantities:
$$\sup_{\theta\in\Theta}\vert \nabla\lambda_{1,t}(\theta)-\nabla\overline{\lambda}_{1,t}(\theta)\vert,\quad \sup_{\theta\in\Theta}\vert \lambda_{1,t}(\theta)-\overline{\lambda}_{1,t}(\theta)\vert\times \sup_{\theta\in\Theta}\Vert \nabla \overline{\lambda}_{1,t}(\theta)\Vert,$$
$$\exp\left(\sup_{\theta\in\Theta}\vert\lambda_{2,t}(\theta)\vert\right)\times \sup_{\theta\in\Theta}\Vert \nabla \overline{\lambda}_{1,t}(\theta)\Vert\times \exp\left(\sup_{\theta\in\Theta}\vert \lambda_{2,t}(\theta)\vert+\sup_{\theta\in\Theta}\vert \overline{\lambda}_{2,t}(\theta)\vert\right)\times \sup_{\theta\in\Theta}\vert \lambda_{2,t}(\theta)\overline{\lambda}_{2,t}(\theta)\vert.$$
Using Lemma \ref{lem::approxlatent} and the integrability conditions of Lemma \ref{lem::momlatent} as well as the existence of all the exponential moments for $\sup_{\theta\in\Theta}\vert \lambda_{2,t}(\theta)\vert$ already justified above,
we easily get the result.$\square$

\paragraph{Proof of Proposition \ref{pp::consistencyGAIN}}

We check the assumptions of Theorem \ref{consistency}. First, note that Assumption {\bf A7} follows directly form our assumptions and Lemma \ref{lem::identifiability}. Moreover, checking Assumption {\bf A4} is straightforward and follows form standard arguments. 
We then check {\bf A5} and define $h_t(s)=h_{1,Y_{1,t}}(s_1)+h_{1,Y_{2,t}}(s_2)$.
We have
\begin{eqnarray*}
\sup_\theta |h_t(\lambda_t(\theta))| & \leq &  d_-^{-1} \left(Y_{1,t}^2 + \sup_\theta |\lambda_{1,t}(\theta)| + 1\right) + d_-^{-1}  Y_{2,t} \\
 & & +  \sup_\theta |\lambda_{2,t}(\theta)| + Y_{2,t}  \left(\log(d_-) + \log\left( 1 + \sup_\theta |\lambda_{2,t}(\theta)| \right)\right)
\end{eqnarray*}
Since $\rho(B_0 + A_0) < 1$ and $\E(\epsilon^2) = 1$, one can found $r < \delta$ small enough such that  $\rho(B_0 + A_0 \mathrm{diag}(\E^{1/(1+r)}[\epsilon ^ {2(1+r)}], 1 ) < 1$. Then, from Proposition \ref{pp::momentGAIN}, we have for $\delta$ small enough, $\E[|\overline{Y}_0|^{1+\delta}] < \infty$ and Lemma \ref{lem::momlatent} yields  to  $\E[(sup_\theta|\lambda_0(\theta)|_1)^{1+\delta}] <  \infty$.  Hence  $\E\left[ \log^{1 + 1/\delta}\left( 1 + \sup_\theta |\lambda_{2,t}(\theta)| \right)\right] < \infty$ and consequently    
$\E\left[ Y_{2,t} \log \left( 1 + \sup_\theta |\lambda_{2,t}(\theta)| \right)\right] < \infty.$ It follows that $\E[\sup_\theta |h_t(\lambda_t(\theta))| ] < \infty$. We then conclude that $\E(\sup_{\theta \in \Theta}|h_t(\lambda_t(\theta))|) < \infty$ and {\bf A5} follows.
Finally, we check {\bf A6}.
Using the Lipschitz property of the function $h_t$ and Lemma \ref{lem::approxlatent}, we have
\begin{eqnarray*}
|h_t(\overline{\lambda}_t(\theta)) - h_t(\lambda_t(\theta)) | &  \leq  & 2 \left(d_- ^{-2} Y_{1,t}^2 + Y_{2,t} + d_- ^{-1} + 1\right)\vert \lambda_t(\theta) - \overline{\lambda}_t(\theta)\vert_1 \\
 & \leq & C (\sup_{\theta\in\Theta}|\lambda_0(\theta) |_1 + \vert \overline{\lambda}_0\vert_1) \left(d_- ^{-2} Y_{1,t}^2 + Y_{2,t} + d_- ^{-1} + 1\right) \tau^t
\end{eqnarray*}
for some constants $C>0, \tau \in (0,1).$ Since the logarithmic moment of $(\sup_{\theta\in\Theta}|\lambda_0(\theta) |_1 + |\overline{\lambda}_0 |_1) \left(d_- ^{-2} Y_{1,t}^2 + Y_{2,t} + d_- ^{-1} + 1\right)$ is finite, then $\sum_{t \geq 1} \sup_{\theta \in \Theta}  |h_t(\overline{\lambda}_t(\theta)) - h_t(\lambda_t(\theta)) | < \infty$ and consequently, almost surely, as $n$ tends to infinity,  
$$n^{-1}\sum_{t = 1}^n \sup_{\theta \in \Theta} |h_t(\lambda_t(\theta)) - h_t(\overline{\lambda}_t(\theta))| \rightarrow 0. \square$$ 

\subsubsection{Proof of Proposition \ref{pp::asnormalityGAIN}} 

We check the assumptions of Theorem \ref{asnormality}, in particular {\bf A8-A12}. Note that {\bf A12} follows directly from Lemma \ref{lem::latentdevfree}.
Moreover, checking {\bf A8} is straightforward and then omitted.
We next check {\bf A9} and {\bf A10}.
Due to the specific from of $h_{1,y}$ and $h_{2,y}$, we only have to check the integrability of the following random variables.
$$\lambda_{1,0}(\theta_0)\sup_{\theta\in\Theta}\frac{\Vert \nabla \lambda_{1,0}(\theta)\Vert^2}{\lambda_{1,0}(\theta)^2},\quad \sup_{\theta\in\Theta}\frac{\Vert \nabla \lambda_{1,0}(\theta)\Vert^2}{\lambda_{1,0}(\theta)},\quad \lambda_{1,0}(\theta_0)\sup_{\theta\in\Theta}\frac{\Vert \nabla^{(2)}\lambda_{1,0}(\theta)\Vert}{\lambda_{1,0}(\theta)^3},\quad \sup_{\theta\in\Theta}\frac{\Vert\nabla^{(2)}\lambda_{1,0}(\theta)\Vert}{\lambda_{1,0}(\theta)^2},$$
as well as 
$$\sup_{\theta\in\Theta}\Vert \nabla \lambda_{2,0}(\theta)\Vert^2,\quad \lambda_{2,0}(\theta_0)\left(\sup_{\theta\in\Theta}\frac{\Vert \nabla \lambda_{2,0}(\theta)\Vert^2}{\lambda_{2,0}(\theta)}+\sup_{\theta\in\Theta}\frac{\Vert \nabla^{(2)} \lambda_{2,0}(\theta)\Vert}{\lambda_{2,0}(\theta)^2}\right).$$
All these integrability conditions follows from Lemma \ref{lem::smoments}, Lemma \ref{lem::momlatent} and Proposition \ref{pp::momentGAIN} with $r=1+\delta$.

Finally, we check {\bf A11}.
To this end, it is sufficient to show that $\sum_{t=1}^{\infty}\sup_{\theta\in\Theta}\zeta_t(\theta)<\infty$, when $\zeta_t(\theta)$ is one of the following quantities.
$$\Vert \nabla \lambda_{1,t}(\theta)-\nabla \overline{\lambda}_{1,t}(\theta)\Vert\times\lambda_{1,t}(\theta_0),\quad \left\vert \lambda_{1,t}(\theta)-\overline{\lambda}_{1,t}(\theta_0)\right\vert\times \Vert \nabla \overline{\lambda}_{1,t}(\theta)\Vert, \quad \frac{\Vert \nabla \overline{\lambda}_{1,t}(\theta)\Vert}{\lambda_{1,t}(\theta)}\times \vert \lambda_{1,t}(\theta)-\overline{\lambda}_{1,t}(\theta)\vert\times \lambda_{1,t}(\theta_0),$$
$$\Vert \nabla \lambda_{2,t}(\theta)-\nabla \overline{\lambda}_{2,t}(\theta)\Vert\times\lambda_{2,t}(\theta_0),\quad \frac{\Vert \nabla\overline{\lambda}_{2,t}(\theta)\Vert}{\lambda_{2,t}(\theta)}\times \vert \lambda_{2,t}(\theta)-\overline{\lambda}_{2,t}(\theta)\vert\times \lambda_{2,t}(\theta_0).$$
The result follows from the approximation results given in Lemma \ref{lem::approxlatent} and the integrability conditions given by Lemmas \ref{lem::smoments}, \ref{lem::momlatent} and Proposition \ref{pp::momentGAIN}.$\square$

\subsubsection{Proof of Proposition \ref{consistCop}}
From {\bf R1-R2} and the continuity assumption on $f_1$, we have a uniform law of large numbers. In particular, 
$$\sup_{R\in\Gamma}\left\vert \frac{1}{n}\sum_{t=1}^n f_t\left(\hat{\theta},R\right)-\overline{f}\left(\theta_0,R\right)\right\vert\rightarrow 0\mbox{ a.s}.$$
Next, 
$$\frac{1}{n}\sum_{t=1}^n f_t\left(\hat{\theta},\hat{R}\right)\geq \frac{1}{n}\sum_{t=1}^n f_t\left(\hat{\theta},R_0\right).$$
Assume that $\widetilde{R}$ is a cluster point of the sequence $\left(\hat{R}(\omega)\right)_n$ for an $\omega$ such that the previous uniform convergence holds true. Taking the limit in the previous equality, we get 
$$\overline{f}\left(\theta_0,\widetilde{R}\right)\geq \overline{f}\left(\theta_0,R_0\right).$$
Then $\widetilde{R}\in \mathcal{I}_0$. Hence, $d\left(\hat{R},\mathcal{I}_0\right)\rightarrow 0$ a.s. 

We next study the convergence in total variation distance. To this end, we give another description of the set $\mathcal{I}_0$. Denoting by 
$$KL_{\lambda}\left(R_0,R\right)=\int \log\left(\frac{p_{R}(y\vert \lambda)}{p_{R_0}(y\vert\lambda)}\right) p_{R_0}(y\vert\lambda)\mu(dy)$$
the Kullback-Leibler divergence between $P_{R}\left(\cdot\vert \lambda\right)$ and $P_{R_0}\left(\cdot\vert \lambda\right)$ which is a non-negative quantity,
we have 
$$\overline{f}\left(R,\theta_0\right)-\overline{f}\left(R_0,\theta_0\right)=\E\left[KL_{\lambda_0}\left(R_0,R\right)\right].$$
Hence, if $R\in\mathcal{I}_0$, we have $KL_{\lambda}\left(R_0,R\right)=0$ for every $\lambda$ in an event of $\P_{\lambda_0}-$probability one.
From {\bf R4}, $R\mapsto KL_{\lambda}\left(R_0,R\right)$ is continuous for every $\lambda$. We then deduce the existence of a measurable set 
$\Lambda$ such that $\P_{\lambda_0}\left(\Lambda\right)=1$ and for every $\lambda\in\Lambda$ and $R\in\mathcal{I}_0$, 
$KL_{\lambda}\left(R_0,R\right)=0$. Now let us show that almost surely,
\begin{equation}\label{tvcov}
KL_{\lambda}\left(R_0,\hat{R}\right)\rightarrow 0,\quad\lambda\in\Lambda.
\end{equation}
Let $\lambda\in\Lambda$. Since any cluster point $\widetilde{R}$ of a sequence $\hat{R}(\omega)$ is in $\mathcal{I}_0$, we have $KL_{\lambda}\left(R_0,\widetilde{R}\right)=0$ and then 
(\ref{tvcov}) follows. Using Pinsker's inequality (the total variation distance is bounded by the square root of one half of the Kullback-Leibler divergence), we also get the convergence in total variation distance.$\square$

\subsubsection{Proof of Theorem \ref{consistGauss}}
If $R$ is in a compact set $\Gamma$, there exist some positive real numbers $\alpha_0,\beta_0,\alpha_1,\beta_1$ such that for any $R\in\Gamma$,
\begin{equation}\label{bounds}
\alpha_0 \exp\left(-\frac{1}{2}\left(\beta_0-1\right)\Phi^{-1}(u)'\Phi^{-1}(u)\right)\leq c_{R}(u)\leq \alpha_1\exp\left(-\frac{1}{2}\left(\beta_1-1\right)\Phi^{-1}(u)'\Phi^{-1}(u)\right).
\end{equation}

In what follows, we will derive a lower and an upper bound for the integral
$$I=\int_{d_{\ell+1}}^{e_{\ell+1}}\cdots\int_{d_k}^{e_k}c_{R}\left(u_1,\ldots,u_k\right)du_{\ell+1}\cdots du_k,$$
where $0\leq d_i<e_i\leq 1$ for $\ell+1\leq i\leq k$.

\paragraph{Upper bound for $I$}

Getting an upper bound for $I$ is straightforward. From (\ref{bounds}), we have 
$$I\leq I'\alpha_1\prod_{i=\ell+1}^k \int_0^1 \exp\left(-\frac{1}{2}(\beta_1-1)\Phi^{-1}(u_i)^2\right)du_i=\alpha_1 J_1^{k-\ell},$$
where $I'=\prod_{i=1}^{\ell}\exp\left(-\frac{1}{2}(\beta_1-1)\Phi^{-1}(u_i)^2\right)$ and (after a change of variable $x=\Phi^{-1}(u_i)$),  
$$J_1=\int_{-\infty}^{\infty}\exp\left(-\frac{1}{2}\beta_1 x^2\right)dx=\sqrt{2\pi \beta_1^{-1}}.$$

\paragraph{Lower bound for $I$}

Setting
$$I^{''}=\prod_{i=1}^{\ell}\exp\left(-\frac{1}{2}(\beta_0-1)\Phi^{-1}(u_i)\right)^2$$
and using again (\ref{bounds}), we have 
$$I\geq I^{''}\alpha_0 \prod_{i=\ell+1}^k \int_{d_i}^{e_i}\exp\left(-\frac{1}{2}(\beta_0-1)\Phi^{-1}(u_i)^2\right)d u_i.$$
It is then necessary to get a lower bound for 
$$J=\int_d^e \exp\left(-\frac{1}{2}(\beta_0-1)\Phi^{-1}(u)^2\right)du,$$
for some real numbers $0\leq d <e\leq 1$. We consider several cases.
\begin{enumerate}
\item
Suppose first that $0<d<e<1$. In this case, we have 
$$J\geq (e-d)\exp\left(-\frac{1}{2}(\beta_0-1)\Phi^{-1}(e)^2\right)\exp\left(-\frac{1}{2}(\beta_0-1)\Phi^{-1}(d)^2\right).$$
\item
Assume now that $d=0$ and $e<1$. In this case 
$$J=\int_{-\infty}^{\Phi^{-1}(e)}\exp\left(-\frac{1}{2}\beta_0 x^2\right)dx=\sqrt{2\pi \beta_0^{-1}}\Phi\left(\sqrt{\beta_0}\Phi^{-1}(e)\right).$$
Using the inequality $\Phi(x)+\Phi(-x)=1$ and the inequality
$$1-\Phi(x)\geq \frac{\exp\left(-\frac{x^2}{2}\right)}{2\sqrt{2\pi}x},\quad x\geq 1,$$
we get for $x\geq 1$, $1-\Phi(x)\geq f \exp(-x^2)$ for a suitable constant $1\geq f>0$ such that $f\leq \Phi(-1)$ and $f\leq\sqrt{2\pi \beta_0^{-1}}$. Then  
$$\Phi(x)\geq \Phi(-1)\mathds{1}_{x\geq -1}+\left(1-\Phi(-x)\right)\mathds{1}_{x<-1}\geq f \exp(-x^2).$$
We then get
$$J\geq f^2\exp\left(-\beta_0\Phi^{-1}(e)^2\right).$$
\item
Assume next that $e=1$ and $d>0$. We also get 
$$\sqrt{2\pi \beta_0^{-1}}\left(1-\Phi\left(\sqrt{\beta_0}\Phi^{-1}(d)\right)\right)\geq f^2 \exp\left(-\beta_0\Phi^{-1}(d)^2\right).$$
\item
Finally if $e=1$ and $d=0$, then $J\geq \sqrt{2\pi \beta_0^{-1}}\geq f\geq f^2$.
\end{enumerate}

We then showed the following result.

\begin{lem}\label{util+}
\begin{enumerate}
\item
There exist some real numbers $f_1$ and $f_2$, not depending on the $d_i'$s and the $e_i'$s such that
$$\log I\leq f_1+f_2\sum_{i=1}^{\ell}\Phi^{-1}(u_i)^2.$$
\item
There exist some real numbers $f'_1,f'_2$ and $f'_3$, not depending on the $d_i'$s and the $e_i'$s such that
$$\log I\geq f'_1+f'_2\sum_{i=1}^{\ell} \Phi^{-1}(u_i)^2+\sum_{i=\ell+1}^k\left[\log(e_i-d_i)\mathds{1}_{0<d_i<e_i<1}+f'_3 \Phi^{-1}(e_i)^2\mathds{1}_{e_i<1}+f'_3\Phi^{-1}(d_i)^2\mathds{1}_{d_i>0}\right].$$
\end{enumerate}

\end{lem}

The next lemma will be also needed.

\begin{lem}\label{gaus}
\begin{enumerate}
\item
There exist $\delta_1>0$ such that for any $u\in (0,1)$,
$$\Phi^{-1}(u)^2\leq \delta_1\left(1-\log(u)-\log(1-u)\right).$$
\item
Let $X$ be a random variable supported on the integers and such that $p_0=\P(X=0)\in (0,1)$. If $F$ denotes the cdf of $X$, we have the bound
$$\E\left[\Phi^{-1}\left(F(X^{-})\right)^2\mathds{1}_{X\geq 1}\right]\leq \delta_2\left(1-\log(p_0)-\log(1-p_0)\right),$$
where $\delta_2$ does not depend on $F$. 
\item
Let $X$ be a random variable supported on the integers and such that $p_k=\P(X=k)\in (0,1)$ for any $k\in \N$. If $F$ denotes the cdf of $X$, we have the bound
$$\E\Phi^{-1}\left(F(X)\right)^2\leq \delta_1\left(1-\log(p_0)+\E\log(1-F(X))\right),$$
where $\delta_3>0$ does not depend on $F$.
\end{enumerate}
\end{lem}

\paragraph{Proof of Lemma \ref{gaus}}

\begin{enumerate}
\item
Since $\Phi^{-1}(u)\sim \sqrt{-2\log(u)}$ when $u\sim 0$ and $\Phi^{-1}(u)\sim \sqrt{-2\log(1-u)}$ when $u\sim 1$, the result is straightforward.
\item
We represent $X$ as $F^{-1}(U)$.
On the event $\{X\geq 1\}$, we have $F(X^{-})\geq p_0$ and 
$$F(X^{-})=\sum_{k=1}^{\infty}F(k-1)\mathds{1}_{F(k-1)<U\leq F(k)}\leq U.$$
Since $\E\Phi^{-1}(U)^2=1$, the result follows from the bound given in the previous point. 
\item
Using the first point of the lemma, it is only necessary to bound $-\E\log(F(X))\leq -\log(p_0)$.$\square$
\end{enumerate}

We now go back to the proof of Proposition \ref{consistGauss}. It is only necessary to check {\bf A14} and {\bf A16}. {\bf A15} easily follows from the standard properties of Kullback-Leibler divergence.
\begin{enumerate}
\item
We first check {\bf A16}. We use the fact that $R\mapsto c_{R}$ is continuous and we apply the dominated convergence theorem.
To this end, we use the upper/lower bound on $I$ given in Lemma \ref{util+} and it is necessary to check the following integrability conditions.
\begin{equation}\label{ultra1}
\int \Phi^{-1}\left(F_{i,\lambda_i}(y)\right)^2p_{i,\lambda_i}(y)dy<\infty,\quad 1\leq i\leq \ell,
\end{equation}
\begin{equation}\label{ultra2}
\int \Phi^{-1}\left(F_{i,\lambda_i}(y)\right)^2 \mathds{1}_{F_{i,\lambda_i}(y)<1} p_{i,\lambda_i}(y)d\mu_i(y)<\infty,\quad \ell+1\leq i\leq k, 
\end{equation}
\begin{equation}\label{ultra3}
\int \Phi^{-1}\left(F_{i,\lambda_i}(y^{-})\right)^2 \mathds{1}_{F_{i,\lambda_i}(y^{-})>0} p_{i,\lambda_i}(y)d\mu_i(y)<\infty,\quad \ell+1\leq i\leq k, 
\end{equation}
\begin{equation}\label{ultra4}
\int -\log\left(p_{i,\lambda_i}(y)\right)p_{i,\lambda_i}(y)d\mu_i(y)<\infty,\quad \ell+1\leq i\leq k. 
\end{equation}
Checking (\ref{ultra1}) is automatic by continuity of $F_{i,\lambda_i}$, since any integral of this form writes as $\E\Phi^{-1}(U)^2=1$ where $U$ is a uniformly distributed over $[0,1]$. 
(\ref{ultra2}) and (\ref{ultra4}) follow from Assumption {\bf G1}. Moreover, it is easy to check that (\ref{ultra3}) is valid either for the Bernoulli distribution or for any distribution with full support $\N$.

\item
We next check {\bf A14}.
From Lemma \ref{util+} and Lemma \ref{gaus}, we only have to check the following integrability conditions.
When $1\leq i\leq\ell$, we have to show that
\begin{equation}\label{conti}
\E\left[\sup_{\theta\in\Theta}\left\{-\log F_{i,\lambda_{i,0}(\theta)}(Y_{i,0})\right\}+\sup_{\theta\in\Theta}\left\{-\log \left(1-F_{i,\lambda_{i,0}(\theta)}(Y_{i,0})\right)\right\}\right]<\infty.
\end{equation}
When $\ell+1\leq i\leq k$, we have to show that
\begin{equation}\label{disc1}
\E\left[\sup_{\theta\in\Theta}\left\{-\log p_i\left(0\vert \lambda_{i,0}(\theta)\right)\right\}+\sup_{\theta\in\Theta}\left\{-\log \left(1-p_i\left(0\vert\lambda_{i,0}(\theta)\right)\right)\right\}\right]<\infty,
\end{equation}
\begin{equation}\label{disc2}
\E\left[\sup_{\theta\in\Theta}\left\{-\log p_i\left(Y_{i,0}\vert \lambda_{i,0}(\theta)\right)\right\}\right]<\infty
\end{equation}
and for count marginal time series,
\begin{equation}\label{disc3}
\E\left[\sup_{\theta\in\Theta}\left\{-\log \left(1-F_{i,\lambda_{i,0}(\theta)}(Y_{i,0})\right)\right\}\right]<\infty.
\end{equation}
These conditions are precisely ensured by Assumptions {\bf G2-G3}.
\end{enumerate}

To end the proof of Theorem \ref{consistGauss}, it is necessary to check that parameter $R$ can be identified, i.e. that $\mathcal{I}_0=\{R_0\}$ where $\mathcal{I}_0$ is defined before the statement of Proposition \ref{consistCop}. From the properties of Kullback-Leibler divergence, it is sufficient to show that if $p_R(\cdot\vert s)=p_{R_0}(\cdot\vert s)$, $\mu-$almost everywhere for some $s\in F_1\times\cdots\times F_k$, than $R=R_0$.
Such identification property is already known in the literature. See for instance \citet{Marbac}, appendix A. For simplicity, we summarize the required arguments, using our notations.
To show this, we first give an expression of the density $p_R(\cdot \vert s)$ which will be simply denoted $p_R(\cdot)$ here.
Moreover, we simply denote by $p_i$ the density $p_i(\cdot\vert s_i)$ and $F_i=F_{i,s_i}$.
In what follows, for any value of $k$, we denote by $\Phi_R$ the Gaussian density with mean $0$ and covariance matrix $R$ and simply by $\phi$ the density of the standard Gaussian distribution on the real line. We have 
$$\frac{p_R(y)}{\prod_{i=1}^{\ell}p_i(y_i)}=\int_{F_{\ell+1}(y_{\ell+1}^{-})}^{F_{\ell+1}(y_{\ell+1})}\cdot \int_{F_k(y_k^{-})}^{F_k(y_k)} 
\frac{\phi_R\left(\Phi^{-1}(F_1(y_1)),\cdot,\Phi^{-1}(F_{\ell}(y_{\ell})),\Phi^{-1}(u_{\ell+1}),\cdot,\Phi^{-1}(u_k)\right)}{\phi_{I_k}\left(\Phi^{-1}(F_1(y_1)),\cdot,\Phi^{-1}(F_{\ell}(y_{\ell})),\Phi^{-1}(u_{\ell+1}),\cdot,\Phi^{-1}(u_k)\right)}du_{\ell+1}\cdot du_k.$$
Suppose that $\Phi_R=\Phi_{R_0}$, $\mu-$a.e.
For $1\leq i<j\leq k$, let $R(i,j)=\begin{pmatrix}1&r_{ij}\\r_{ij}&1\end{pmatrix}$, which is simply the (sub-)correlation matrix for components $i$ and $j$.
Finally, we denote by $p_{R(i,j)}$ the bivariate density corresponding to these components.
We consider three cases.
\begin{enumerate}
\item
Assume first that $1\leq i\leq \ell$ and $\ell+1\leq j\leq k$. In this case, we have 
\begin{eqnarray*}
p_{R(i,j)}(y_i,y_j)&=&p_i(y_i) \int_{\Phi^{-1}(F_j(y_j^{-}))}^{\Phi^{-1}(F_j(y_j))}\frac{\phi_{R_{ij}}\left(\Phi^{-1}(F_i(y_i)),x_j\right)}{\phi_{I_1}\left(\Phi^{-1}(F_i(y_i))\right)}dx_j\\
&=& p_i(y_i)\left\{\Phi\left(\frac{\Phi^{-1}(F_j(y_j))-r_{ij}\Phi^{-1}(F_i(y_i))}{\sqrt{1-r_{ij}^2}}\right)-\Phi\left(\frac{\Phi^{-1}(F_j(y_j^{-}))-r_{ij}\Phi^{-1}(F_i(y_i))}{\sqrt{1-r_{ij}^2}}\right)\right\}.
\end{eqnarray*}
Then if $p_{R(i,j)}=p_{R_0(i,j)}$ almost everywhere, there exists $w\in \R$ such that 
$$\Phi\left(\frac{w-r_{ij}\Phi^{-1}(F_i(y_i))}{\sqrt{1-r_{ij}^2}}\right)=\Phi\left(\frac{w-r_{0ij}\Phi^{-1}(F_i(y_i))}{\sqrt{1-{r_{0ij}}^2}}\right),$$
for almost every value of $y_i$ (with respect to the Lebesgue measure).
Since $\Phi$ is one-to-one and $F_i(y_i)$ can take arbitrary values between $0$ and $1$, it is easily seen that $r_{ij}=r_{0ij}$.
\item
Assume now that $\ell+1\leq i<j\leq k$. In this case, we have
$$p_{R(i,j)}(y_i,y_j)=\int_{\Phi^{-1}(F_i(y_i^{-}))}^{\Phi^{-1}(F_i(y_i))}\int_{\Phi^{-1}(F_j(y_j^{-}))}^{\Phi^{-1}(F_j(y_j))}\phi_{R(i,j)}dxi dx_j.$$
We use the expression,
$$\phi_{R(i,j)}(x_i,x_j)=\phi(x_i)\left(2\pi(1-r_{ij}^2)\right)^{-1/2}\exp\left(-\frac{(x_j-r_{ij}x_i)^2}{2(1-r_{ij}^2)}\right).$$
Whatever the cases (binary or count variables), if $p_{R(i,j)}=p_{R_0(i,j)}$ almost everywhere, there exists two real numbers $w_i$ and $w_j$ such that 
we have the equality $f(r_{ij})=f(r_{0ij})$ with
$$f(r)=\int_{-\infty}^{w_i}\Phi\left(\frac{w_j-r x_i}{\sqrt{1-r^2}}\right)\phi(x_i)dx_i.$$
However, since after some computations, the derivative of $f$ can be written as
$$\dot{f}(r)=-(1+d^2)^{-1}\phi(c-d w_i)\phi(w_i),$$
with $c=(1-r^2)^{-1/2}w_j$ and $d=(1-r^2)^{-1/2}r$, we see that $f$ is decreasing. Hence, $r_{ij}=r_{0ij}$.
\item
Finally, if $1\leq i<j\leq \ell$, we recover the identification problem for continuous margins with
$$p_{R_{ij}}(y_i,y_j)=p_i(y_i)p_j(y_j)\frac{\phi_{R_{ij}}\left(\Phi^{-1}(F_i(y_i)),\Phi^{-1}(F_j(y_j))\right)}{\phi\left(\Phi^{-1}(F_i(y_i))\right)\phi\left(\Phi^{-1}(F_j(y_j))\right)}.$$
Identification of $r_{ij}$ is straightforward in this case.
\end{enumerate}
We then deduce that the set $\mathcal{I}_0$ only contains $R_0$ and the consistency result now follows from Proposition \ref{consistCop}.$\square$

\subsection{Proof of Corollary \ref{contentGAIN}}
We check the assumptions of Theorem \ref{consistGauss}.
When $p_{i,s_i}$ is the Poisson distribution with parameter
$\phi(s_i)$, with $\phi(s_i)=s_i$ or $\phi(s_i)=\exp(s_i)$, it is straightforward to check the first and the third conditions {\bf G1}.
It remains to check the second one. If $X$ follows a Poisson distribution with parameter $\mu$ and denoting by $F$ its cdf, we have
\begin{equation}\label{queue} 
1-F(k)\geq \exp(-\mu)\frac{\mu^{k+1}}{(k+1)!}.
\end{equation}
The previous lower bound follows from a Taylor-Lagrange expansion. We then get 
$$-\log(1-F(k))\leq \log(\mu)-(k+1)\log(\mu)+\sum_{i=1}^{k+1}\log(i)\leq -k\log(\mu)+\frac{k(k+1)}{2}.$$
In the last bound, we have simply used the bound $\log(i)\leq i-1$ for $i\geq 1$.
We then get
$$-\E\log(1-F(X))\leq C_1(1+\mu^2),$$
where $C_1>0$ does not depend on $F$. 

Next we check {\bf G2}. It is easily seen that if $p_{i,s_i}$ is the Poisson distribution with parameter
$s_i$, then the three conditions in {\bf G2} are satisfied as soon as $\E\sup_{\theta\in\Theta}\lambda_{i,0}(\theta)^{1+\delta}<\infty$, which is guaranteed from the assumptions of Proposition \ref{pp::consistencyGAIN}.

Finally, {\bf G3} is satisfied for the GARCH component as soon as $\E\sup_{\theta\in\Theta}\lambda_{i,0}(\theta)<\infty$, which is also automatic under the assumptions of Proposition \ref{pp::consistencyGAIN}.
The result then follows from Theorem \ref{consistGauss}.$\square$

\subsection{Proof of Corollary \ref{contentBIP}}
The proof is similar to that of Corollary \ref{contentGAIN}.
One can show that {\bf G2} is satisfied for the Poissonian component as soon as 
$\E \exp\left((1+\delta)\sup_{\theta\in\Theta}\vert\lambda_{i,0}(\theta)\vert\right)<\infty$ for some $\delta>0$, which is covered by our assumptions.
Note that (\ref{disc30}) is satisfied under the same type of conditions by using the lower bound (\ref{queue}) for the survival function of a Poisson distribution.  

For the binary coordinate with $p_i(1\vert \lambda_i)=F(\lambda_i)$ where $F$ is the logistic cdf. In this case, it is only necessary to check (\ref{disc10}).
The required conditions are satisfied if $\E\sup_{\theta\in\Theta}\vert \lambda_{i,0}(\theta)\vert<\infty$ which is the case under our assumptions.
The other conditions in {\bf G1} are trivial to show or has been discussed in the proof of Corollary \ref{contentGAIN}. The result then follows from Theorem \ref{consistGauss}.$\square$

\section{Appendix}\label{appendix}

\subsection{Two useful lemmas}

\begin{lem}
\label{lem::poism}
Let $\lambda>0$ and $X_{\lambda}$ $\mathrm{Poisson}$ variable  with parameter $\lambda$.   Then, $\forall r \geq 1$ and any $\delta\in (0,1)$,  there exists $b_{r,\delta}$, not depending on $\lambda$ and such that
$$
\|X_{\lambda}\|_r \leq (1+\delta)\lambda+ b_{r,\delta}.
$$
\end{lem}
\paragraph{Proof of lemma \ref{lem::poism}}
We have the equality $\mathbb{E}(X_{\lambda}^r) = \sum_{i = 1}^r \lambda^i \left\{\begin{array}{c}
   r  \\
    i 
\end{array} \right\}  $ with  $\left\{\begin{array}{c}
    r  \\
    i 
\end{array} \right\}$ are the  Sterling's numbers of second kind. 
See for instance \cite{johnson2005univariate}.

Then 
$$
\mathbb{E}(X_{\lambda}^r)  = \lambda^r + \sum_{i = 1}^{r-1} \lambda^i \left\{\begin{array}{c}
    r  \\
    i 
\end{array}\right\} 
\leq  \lambda^r + C_r(\lambda + \lambda^{r-1}),
$$
where $C_r>0$ only depends on $r$.
But, we can notice that,  for any $\delta>0$, there exists $\exists \overline{b}_{\delta, r}>0$ such that for all $x\geq 0$ : $x + x^{r-1} \leq \delta' x^r + \overline{b}_{\delta, r}$ with $\delta'=\frac{(1+\delta)^r-1}{C_r}$. Then $\mathbb{E}(X^r) \leq (1+C_r\delta')\lambda^r + C_r \overline{b}_{\delta, r}$. 
Therefore $\|X\|_r \leq (1+C_r\delta')^{1/r} \lambda + C_r^{1/r}\overline{b}_{\delta, r}^{1/r}$.
Setting $b_{\delta,r}=C_r^{1/r}\overline{b}_{\delta, r}^{1/r}$, we get the result.$\square$

\begin{lem}
\label{lem::bornerayon} Let $B(\theta)$ be a matrix with entries depending continuously on a parameter $\theta \in \Theta$ and $\Theta$ is a compact set of $\R^d, d \in \N^*$. Suppose that $\rho\left(B(\theta)\right)<1$ for any $\theta\in\Theta$. There then exist $C>0$ and $\tau \in (0,1)$ such that for all integer $j\geq 1$, $\sup_{\theta\in\Theta} |B(\theta)^j|_1 \leq C \tau^j$. 
\end{lem}

\paragraph{Proof of Lemma \ref{lem::bornerayon}}
Let $\Vert\cdot\Vert$ be an arbitrary norm on $\R^d$.
For any   $\theta \in \Theta$,  from Gelfand's formula  $\lim_{n \to \infty} |B(\theta)^n|_1^{1/n} = \rho(B(\theta))$.
Therefore there exists $\rho_\theta \in (0,1)$ and  $n_\theta \in \mathbb{N}$ such that  $|B(\theta)^{n_\theta}|_1 < \rho_\theta$. By continuity of the function $\theta \mapsto B(\theta)$, we can found  $\epsilon_\theta$ such that  $\forall \overline{\theta} \in \mathcal{B}(\theta, \epsilon_\theta) = \{\eta \in \Theta :  \|\eta - \theta\| < \epsilon_\theta\}$ , $|B^{n_0}(\overline{\theta})|_1 < \rho_\theta$. By compactness of  $\Theta$ and  Borel-Lebesgue property, $\Theta \subset \bigcup_{i = 1}^N\mathcal{B}(\theta_i, \epsilon_{\theta_i})$ for $\theta_1, \ldots, \theta_N \in \Theta$. Let us set  $\overline{\rho} = \max_{1\leq i \leq N} \rho_{\theta_i} \in (0,1)$ and $n_0 = n_{\theta_1} \times \cdots \times n_{\theta_N}$, it follows that  
$$
\sup_{\theta \in \Theta}|B(\theta)^{n_0}|_1 \leq \overline{\rho}.
$$
If  $n \geq n_0, n = k n_0 + r, k \geq 1, r \in \{0, \ldots, n_0-1\},$ we will set $\overline{C} = \max_{\theta \in \Theta}\left(|B(\theta)|_1 + 1\right)^{n_0},$ and we obtain $$\sup_{\theta \in \Theta}|B(\theta)^n|_1 \leq \overline{C}\overline{\rho}^k = \overline{C}\overline{\rho}^{\lfloor\frac{n}{n_0}\rfloor} \leq \overline{C}\overline{\rho}^{-1}\left(\overline{\rho}^{\frac{1}{n_0}}\right)^n := C \rho^n.\square$$

\subsubsection{Sufficient conditions for finiteness of moments}

The following result gives some sufficient conditions for existence of some moments for the stationary solution of (\ref{eq::model1}). For  a random vector $Z=\left(Z_1,\ldots,Z_d\right), d \in \N^*$, we define $\left\| Z \right\|_{t-1, r, vec} := (\E^{1/r}\left[\vert Z_1\vert^r \vert \mathcal{F}_{t-1}\right], \ldots,  \E^{1/r}\left[\vert Z_d\vert^r \vert \mathcal{F}_{t-1} \right])'$. 

\begin{lem}
\label{lem::momFonc} Assume that Assumptions {\bf A1-A3} hold true.  If in addition,
for an integer $\overline{k}>1$, there exists a vector $\phi:=\left(\phi_1,\ldots,\phi_{\overline{k}}\right)$ of nonnegative continuous functions, a real number $r\geq 1$, a matrix $D\in \mathcal{M}_{\overline{k}}$ with nonnegative elements such that $\rho\left(D\right)<1$ and a $\left(\mathcal{F}_t\right)_{t\in\Z}-$adapted and stationary process $(c_t)_{t\in\Z}$, taking values in  $\R_+^{\overline{k}}$ such that for $s\in R^k$ and $t\in\Z$,
$$
\left\|\phi\left(g(s, F_{s}^{-1}(U_t), X_t)\right) \right\|_{t-1,r, vec} \preccurlyeq  c_{t-1} +  D \phi(s).
$$
Then $\E\left[\vert \phi(\lambda_0)\vert_1^r\right]<\infty$ provided that $\E^{1/r}(|c_0|_1^r) < \infty$,
\end{lem}

\paragraph{Proof of Lemma \ref{lem::momFonc}}
Setting $f_t(s)=g\left(s,F_s^{-1}(U_{t-1}),X_{t-1}\right)$, as in the proof of Theorem \ref{th::stabilityLatent}, we have $\lambda_t=f_t\left(\lambda_{t-1}\right)$ and Theorem $2$ in \citet{DT}
ensures that $\lambda_t=\lim_{m\rightarrow \infty}f_{t-m}^t(s)$ a.s. for any value of $s$. Here $f_{t-m}^t=f_t\circ f_{t-1}\circ\cdots\circ f_{t-m+1}$. 
Since $f_{t-m}^{t-1}$ is measurable with respect to $\mathcal{F}_{t-2}$, we have from our assumption, 
\begin{eqnarray*}
\left\|\phi\left(f_{t-m}^t(s)\right)\right\|_{t-2,r, vec} & = &\left\|\phi\left(g(f_{t-m}^{t-1}(s) , F_{f_{t-m}^{t-1}(s)}^{-1}(U_{t-1}), X_{t-1})\right) \right\|_{t-2,r, vec} \\
 & \preccurlyeq &   c_{t-2} +  D \phi\left(f_{t-m}^{t-1}\right).
\end{eqnarray*}
From the triangular inequality, we get $\Vert \phi\left(f_{t-m}^t(s)\right)\Vert_{r,vec}\preccurlyeq \Vert c_{t-2}\Vert_{r,vec}+D \Vert\phi\left(f_{t-m}^{t-1}(s)\right)\Vert_{r,vec}$.
Setting $f=\Vert c_0\Vert_{r,vec}$, $h_m(s)=\Vert \phi\left(f_{-m}^0(s)\right)\Vert_{r,vec}$ and using stationarity, we get 
$$h_m(s)\preccurlyeq f+Dh_{m-1}(s)\preccurlyeq \sum_{i=0}^{m-1}D^i f+D^m \phi(s).$$
Letting $m\rightarrow \infty$, the condition $\rho(D)<1$ and Fatou's lemma leads to the result.$\square$

\subsection{Approximation results for linear latent processes}

In this section, we suppose that :
\begin{equation}
\label{eq::linearlambda2}
   ~  \lambda_t(\theta) = d + B \lambda_{t-1}(\theta) + A \overline{Y}_{t-1} + \Gamma X_{t-1}, ~t\in \Z \text{ and }  \theta \in \Theta,   
\end{equation}
The approximate latent process is then define as:
\begin{equation}
\label{eq::linlambapprox}
  ~   \overline{\lambda}_0(\theta) = \overline{\lambda}_0~ ; ~ \overline{\lambda}_t(\theta) = d + B \overline{\lambda}_{t-1}(\theta) + A \overline{Y}_{t-1} + \Gamma X_{t-1}, ~t > 0 \text{ and } \theta \in \Theta.
\end{equation}
The approximate latent process is initialized by a given deterministic vector $\tilde{\lambda}_0$. 
We introduce the following partial derivatives operators :
$$
 \partial_i = \frac{\partial}{\partial \theta_i} ~ \text{ and } \partial_{ij} = \frac{\partial^2}{\partial \theta_i \partial \theta_j}
$$ 
where $\theta_i$ stands for the $i$-th component of the parameters vector $\theta.$
{\bf In the whole subsection, we assume that the process $\left((\overline{Y}_t,X_t)\right)_{t\in\Z}$ is stationary, $\Theta$ is a compact set and for any $\theta\in\Theta$, $\rho(B)<1$}.

\begin{lem}[Moments of latent process]
\label{lem::momlatent}
Suppose that there exists $r\geq 1$ such that $\E(|\overline{Y}_0|_1^r) < \infty$ and $\E(|X_0|_1^r) < \infty$. Then the mapping $\theta \mapsto \lambda_0(\theta)$ is almost surely two times continuously differentiable. Moreover, 
\begin{enumerate}
    \item for $l = 1, \ldots, k,$ $\E[(\sup_\theta  |\lambda_{l,0}(\theta)|) ^r]< \infty$
    \item for $l = 1, \ldots, k, i = 1, \ldots, Q,$ $\E[(\sup_\theta |\partial_i \lambda_{l,0}(\theta)|)^r] < \infty$
    \item for $l = 1, \ldots, k, i,j = 1, \ldots, Q,$ $\E[(\sup_\theta |\partial_{ij} \lambda_{l,0}(\theta)|)^r] < \infty$
\end{enumerate}
\end{lem}
\paragraph{Proof of Lemma \ref{lem::momlatent}}
First note that $\forall \theta,~ \lambda_t(\theta)=\sum_{j \geq 0} B^j\left(d + A \overline{Y}_{t-j-1}+\Gamma X_{t-j-1}\right)$ is well defined and infinitely differentiable since $\sup_{\theta\in \Theta} \rho(B) < 1$. 
Moreover, from Lemma \ref{lem::bornerayon}, there exists $C>0$ and $\tau\in(0,1)$ such that for any $\theta\in\Theta$ and any integer $j\geq 1$, $\vert B^j\vert_1\leq C \tau^j$. Setting $D=\sup_{\theta\in\Theta}\left(\vert d\vert_1+\vert A\vert_1+\vert \Gamma\vert_1\right)$, we get 
$$\E^{1/r}[(\sup_\theta  |\lambda_t(\theta)|_1) ^r]\leq CD\sum_{j\geq 0}\tau^j\left(\E^{1/r}\left(\vert \overline{Y}_{t-j-1}\vert_1^r\right)+\E^{1/r}\left(\vert X_{t-j-1}\vert^r_1\right)\right)$$
which is finite by stationarity and existence of the moment of order $r$.
Next, for all possible indices, $i$ and $j$, the partial derivatives of the latent process are  given by :
\begin{eqnarray}
\label{eq::devpremier}
\dpartial{\lambda_t(\theta)}{d_i} & = & \iota_i + B \dpartial{\lambda_{t-1}(\theta)}{d_i} = \sum_{j \geq 0}B^j \iota_i = (I-B)^{-1} \iota_i,~ i = 1, \ldots, k; \nonumber \\
\dpartial{\lambda_t(\theta)}{A(i,j)} &  =  & E(i,j) \overline{Y}_{t-1} + B \dpartial{\lambda_{t-1}(\theta)}{A(i,j)} = \sum_{l \geq 0} B^l  E(i,j) \overline{Y}_{t-l-1},~ i,j =  1, \ldots, k; \\
\dpartial{\lambda_t(\theta)}{\Gamma(i,j)} & =  & \sum_{l \geq 0} B^l  G(i,j)  X_{t-l-1},~ i =  1, \ldots, k, ~j = 1, \ldots, m ; \nonumber\\
\dpartial{\lambda_t(\theta)}{B(i,j)} &  =  & E(i,j) \lambda_{t-1}(\theta) + B \dpartial{\lambda_{t-1}(\theta)}{B(i,j)} = \sum_{l \geq 0} B^l  E(i,j) \lambda_{t-l-1}(\theta), ~  i,j =  1, \ldots, k \nonumber  
\end{eqnarray}
where $\iota_i, i = 1, \ldots, k$ is the vector of $\{0,1\}^k$ with $1$ at the position $i$  and $0$ elsewhere, 
$E(i,j),  i, j =  1, \ldots, k$ is the $k \times k$ matrix   with $1$ at the position $(i,j)$  and $0$ elsewhere, 
$G(i,j),  i =  1, \ldots, k, j = 1, \ldots, m$ is the $k \times m$ matrix   with $1$ at the position $(i,j)$  and $0$ elsewhere and $d_i$ is the $i-$th element of vector $d.$ 

For all possible indices $i,j,l,v$, the second-order partial derivatives of latent process are given by : 
\begin{eqnarray}
\label{eq::devsecond}
\ddpartial{\lambda_t(\theta)}{d_i}{d_j} & =  & \ddpartial{\lambda_t(\theta)}{d_j}{d_i}=0 \nonumber \\
\ddpartial{\lambda_t(\theta)}{A(i,j)}{A(u,l)} &  =  &  \ddpartial{\lambda_t(\theta)}{\Gamma(i,j)}{\Gamma(u,l)} =0  \nonumber \\ 
\ddpartial{\lambda_t(\theta)}{d_i}{A(j,u)}  & = &  \ddpartial{\lambda_t(\theta)}{A(j,u)}{d_i} = \ddpartial{\lambda_t(\theta)}{A(i,j)}{\Gamma(u,l)} = \ddpartial{\lambda_t(\theta)}{\Gamma(u,l)}{A(i,j)} = 
\ddpartial{\lambda_t(\theta)}{\Gamma(j,u)}{d_i} = \ddpartial{\lambda_t(\theta)}{d_i}{\Gamma(j,u)} = 0  \nonumber \\ 
\ddpartial{\lambda_t(\theta)}{d_l}{B(i,j)} & = &  \ddpartial{\lambda_t(\theta)}{B(i,j)}{d_l} = E(i,j) \dpartial{\lambda_{t-1}(\theta)}{d_l} + B \ddpartial{\lambda_{t-1}(\theta)}{B(i,j)}{d_l}  = \sum_{u \geq 0} B^u E(i,j) \dpartial{\lambda_{t-u-1}(\theta)}{d_l} \\
\ddpartial{\lambda_t(\theta)}{A(l,v)}{B(i,j)} & = &  \ddpartial{\lambda_t(\theta)}{B(i,j)}{A(l,v)} = E(i,j) \dpartial{\lambda_{t-1}(\theta)}{A(l,v)} + B \ddpartial{\lambda_{t-1}(\theta)}{B(i,j)}{A(l,v)}  = \sum_{u \geq 0} B^u E(i,j) \dpartial{\lambda_{t-u-1}(\theta)}{A(l,v)} \nonumber \\
\ddpartial{\lambda_t(\theta)}{\Gamma(l,v)}{B(i,j)} & = &  \ddpartial{\lambda_t(\theta)}{B(i,j)}{\Gamma(l,v)} = E(i,j) \dpartial{\lambda_{t-1}(\theta)}{\Gamma(l,v)} + B \ddpartial{\lambda_{t-1}(\theta)}{B(i,j)}{\Gamma(l,v)}  = \sum_{u \geq 0} B^u E(i,j) \dpartial{\lambda_{t-u-1}(\theta)}{\Gamma(l,v)}  \nonumber \\ 
\ddpartial{\lambda_t(\theta)}{B(l,v)}{B(i,j)} & = &  \ddpartial{\lambda_t(\theta)}{B(i,j)}{B(l,v)} =  \sum_{u \geq 0} B^u \left(E(i,j) \dpartial{\lambda_{t-u-1}(\theta)}{B(l,v)} + E(l,v) \dpartial{\lambda_{t-u-1}(\theta)}{B(i,j)}\right) \nonumber
\end{eqnarray}

Straightforwardly, from Lemma  \ref{lem::bornerayon}, the compactness of $\Theta$ and Minkowski's inequality, all these partial derivatives have  the same  polynomial moments than $\overline{Y}_0$ and $X_0$.$\square$

\begin{lem}[Approximation of the derivatives of the latent process]
\label{lem::approxlatent}
There exist $\tau \in (0,1)$ and $C>0$ such that 
\begin{enumerate}
    \item $\sup_\theta |\lambda_t(\theta) - \tilde{\lambda}_t(\theta)|_1 < C \tau^t (|\lambda_0 |_1 + |\tilde{\lambda}_0 |_1)$
    \item for $ i = 1, \ldots, Q,$ $\sup_\theta |\partial_i \lambda_t(\theta) - \partial_i \tilde{\lambda}_t(\theta)|_1 < C^2 t  \tau^{t-1} (\sup_{\theta\in\Theta}|\lambda_0(\theta)|_1 + |\tilde{\lambda}_0 |_1)+C\tau^t\sup_{\theta\in\Theta}\left\vert \frac{\partial \lambda_0}{\partial B_{i,j}}(\theta)\right\vert.$
\end{enumerate}
  
\end{lem}

\paragraph{Proof of Lemma \ref{lem::approxlatent}}
 One can notice that 
$$
|\lambda_t(\theta) - \overline{\lambda}_t(\theta)|_{vec} \preccurlyeq |B^t|_{vec} |\lambda_0(\theta)  - \overline{\lambda}_{0} |_{vec}
$$
and the first result follows from Lemma \ref{lem::bornerayon}, 1.
For $i,j =  1, \ldots, k,$ we can write  $\dpartial{\lambda_t(\theta)}{B(i,j)}$ as 
$$
\dpartial{\lambda_t(\theta)}{B(i,j)} = \sum_{l = 1}^{t-1} B^l E(i,j) \lambda_{t-l-1}(\theta) + B^t \dpartial{\lambda_0(\theta)}{B(i,j)} \text{ and } \dpartial{\overline{\lambda}_t(\theta)}{B(i,j)} = \sum_{l = 1}^{t-1} B^l E(i,j) \overline{\lambda}_{t-l-1}(\theta) + B^t \dpartial{\overline{\lambda}_0}{B(i,j)}.
$$
And then  
 \begin{eqnarray*}
 \sup_\theta \left\vert \dpartial{\lambda_t(\theta)}{B(i,j)} - \dpartial{\overline{\lambda}_t(\theta)}{B(i,j)} \right\vert_1 & \leq&   \sum_{l = 1}^{t-1} \sup_\theta |B^l|_1 |E(i,j)|_1 \vert \lambda_{t-l-1}(\theta) - \overline{\lambda}_{t-l-1}(\theta)\vert_1+\vert B^t\vert_1\sup_{\theta\in\Theta}\left\vert\dpartial{\lambda_0(\theta)}{B(i,j)}\right\vert\\
 & \leq & tC^2\tau^{t-1}\left(|\lambda_{0}|_1 + |\overline{\lambda}_{0}|_1\right)+C\tau^t\sup_{\theta\in\Theta}\left\vert \frac{\partial \lambda_0(\theta)}{\partial B_{i,j}}\right\vert.\\
 \end{eqnarray*}
The control of the difference between the other partial derivatives is similar.$\square$

\begin{lem}
\label{lem::smoments}
Suppose that all the parameters in \eqref{eq::linearlambda2} are positives and that the processes $(X_t)_{t \in \Z}$ and $(\overline{Y}_t)_{t \in \Z}$ take nonnegative values. Suppose furthermore that there exists some $\delta \in (0,1)$ such that $\E\left(\left\vert\overline{Y}_0\right\vert_1^{\delta}\right) < \infty$ and $\E\left(\left\vert X_0\right\vert_1^{\delta}\right) < \infty$. Then  for any $r\geq 1$,  
$$
\E\left(\sup_\theta \left\vert \frac{1}{\lambda_0(\theta)} \dpartial{\lambda_0(\theta)}{\theta_i}\right\vert_1^r\right) < \infty
\text{ and }
\E\left(\sup_\theta \left\vert \frac{1}{\lambda_0(\theta)} \ddpartial{\lambda_0(\theta)}{\theta_i}{\theta_j}\right\vert_1^r\right) < \infty, i,j = 1, \ldots, Q.
$$
\end{lem}

\paragraph{Proof of Lemma \ref{lem::smoments}}
Note that there exists $d_{-}>0$ such that for any $\theta\in\Theta$, we have $d_i\geq d_{-}$.
Here again, we will  denote by $\iota_\ell, \ell = 1,\ldots, k$ the vector of $\{0,1\}^k$ with $1$ at $\ell-th$ position and $0$ elsewhere.
We also set $a=\min_{\theta\in\Theta}\min_{1\leq i,j\leq k}A(i,j)$, $\gamma=\min_{\theta\in\Theta}\min_{1\leq i,j\leq k}\Gamma(i,j)$ and $b=\min_{\theta\in\Theta}\min_{1\leq i,j\leq k}B(i,j)$ which are positive constant from the positivity assumption and the compactness of $\Theta$. Note that by positivity, all the entries of a matrix of type $B^lA$ are greater than the entries of $aB^lE(i,j)$.
From equations \eqref{eq::devpremier}, we have the bounds
\begin{eqnarray*}
\frac{1}{\lambda_{\ell,t}(\theta)}\dpartial{\lambda_{\ell,t}(\theta)}{d_i} & \leq & \frac{\iota_\ell'(I-B)^{-1}\iota_i}{d_-}; ~\ell, i = 1, \ldots, k,  \\
\frac{1}{\lambda_{\ell,t}(\theta)}\dpartial{\lambda_{\ell,t}(\theta)}{A(i,j)} & \leq & \frac{\iota_\ell' \sum_{l \geq 0} B^l E(i,j) \overline{Y}_{t-l-1}}{a\iota_\ell'\sum_{l \geq 0} B^l E(i,j)\overline{Y}_{t-l-1} } \\ &  &  \leq \frac{1}{a} ; ~\ell, i, j = 1, \ldots, k,   \\
\frac{1}{\lambda_{\ell,t}(\theta)}\dpartial{\lambda_{\ell,t}(\theta)}{\Gamma(i,j)} & \leq & \frac{\iota_\ell' \sum_{l \geq 0} B^l E(i,j) X_{t-l-1}}{\gamma \iota_\ell'\sum_{l \geq 0} B^lE(i,j)X_{t-l-1}} \\ &  &  \leq \frac{1}{\gamma} ; ~\ell, i = 1, \ldots, k,j = 1, \ldots, m.  \\
\end{eqnarray*}
For $t\in \Z$, set $d_t = d + A \overline{Y}_{t-1} + \Gamma X_{t-1}$.  Note that $\dpartial{\lambda_t(\theta)}{B(i,j)} = \sum_{h \geq 1} \sum_{u = 1}^{h} B^{u-1}E(i,j)B^{h-u}d_{t-h}$ and that the entries of $bB^{u-1}E(i,j)B^{h-u}$ are smaller than that of $B^h$. We then obtain
\begin{eqnarray}
\label{eq::devpremiermoment}
\frac{1}{\lambda_{\ell,t}(\theta)}\dpartial{\lambda_{\ell,t}(\theta)}{B(i,j)} & \leq & \sum_{h \geq 1} h \frac{\iota_\ell'  B^h  d_{t-h}}{b\iota_\ell' d +  b  \iota_\ell' B^hd_{t-h}} \\ &  &  \leq \sum_{h \geq 1} h \left(\frac{\iota_\ell'  B^h  d_{t-h}}{b d_{-}}\right)^s ; ~\ell, i, j = 1, \ldots, k,  \nonumber
\end{eqnarray}
for any $s \in (0,1).$ 
From Lemma \ref{lem::bornerayon}, there exist $C>0$ and $\tau\in(0,1)$ such that $\vert B^h\vert_1\leq C \tau^h$ for any positive integer $h$.
We then obtain the bound
$$
\E^{1/r}\left[\sup_{\theta\in\Theta}\left\vert\frac{1}{\lambda_{\ell,t}(\theta)}\dpartial{\lambda_{\ell,t}(\theta)}{B(i,j)}\right\vert^r\right]\leq 
\sum_{h\geq 1}h \frac{C^s\tau^{hs}}{(b^2d_{-})^s}\E^{1/r}\left(\sup_{\theta\in\Theta}\vert d_{t-h}\vert_1^{rs}\right).
$$
Taking $s=\delta/r$ and using the bound $E\left[\left(\sup_\theta|d_1|\right)^\delta \right] \leq \sup_\theta|d|_1^\delta  +  \sup_\theta|A|_1^\delta E\left[|\overline{Y_0}|^\delta \right]+ \sup_\theta|\Gamma|_1^\delta E\left[|X_0|^\delta \right]$, we get the integrability conditions for the first-order partial derivatives.

For the second-order partial derivatives, one can use the expressions \eqref{eq::devsecond} and replace the partial derivatives in the series by the expressions given in \eqref{eq::devpremier}. With more tedious computations, one can use similar arguments as above to get the required integrability conditions. Details are omitted. $\square$

\subsection{Numerical experiments}
\begin{table}[htbp]
	\center 
	\small 
	\caption{Average and Mean Square Errors for the estimators of the BIP model ($n=1000$)}
	\begin{tabular}{|ll|ccccc|ccccc|}
		\cline{3-12}
		\multicolumn{2}{c|}{}  & \multicolumn{5}{c|}{Log INGARCH}  & \multicolumn{5}{c|}{Logit Binary}  \\ \hline
		\multicolumn{2}{|l|}{n = 1000} & \multicolumn{1}{c}{$d_1$} & \multicolumn{1}{c}{$A(1,1)$} & \multicolumn{1}{c}{$A(1,2)$} & \multicolumn{1}{c}{$B(1,1)$} & \multicolumn{1}{c|}{$\Gamma(1,1)$} & \multicolumn{1}{c}{$d_2$} & \multicolumn{1}{c}{$A(2,1)$} & \multicolumn{1}{c}{$A(2,2)$} & \multicolumn{1}{c}{$B(2,2)$} & \multicolumn{1}{c|}{$\Gamma(2,1)$} \\ \hline
		\multicolumn{2}{|c|}{$r$}  & 1 & 0.3 & 0.3 & 0.15 & -0.1 & -1 & 0.4 & -0.6 & 0.2 &  0.1 \\ \hline
		\multicolumn{1}{|l|}{-0.9} & \multicolumn{1}{|c|}{-0.8995 } & 1.0748 & 0.3091 & 0.3047 & 0.1571 & -0.1001 & -1.0251 & 0.4092 & -0.6038 & 0.1958 & 0.1001 \\ 
		& \multicolumn{1}{|c|}{(0.0001)} & 0.0252 & 0.0020 & 0.0008 & 0.0065 & 0.0001 & 0.3877 & 0.0603 & 0.0278 & 0.0331 & 0.0045 \\  \hline
		\multicolumn{1}{|l|}{-0.75 } &  \multicolumn{1}{|c|}{-0.7463} & 1.0649 & 0.3074 & 0.3036 & 0.1634 & -0.1001 & -1.0175 & 0.4058 & -0.6060 & 0.1952 & 0.0995 \\   
		&  \multicolumn{1}{|c|}{ (0.0004)}  & 0.0223 & 0.0016 & 0.0007 & 0.0057 & 0.0001 & 0.2818 & 0.0451 & 0.0229 & 0.0350 & 0.0046 \\  \hline
		\multicolumn{1}{|l|}{-0.6} &   \multicolumn{1}{|c|}{-0.6006  } & 1.0578 & 0.3063 & 0.3032 & 0.1678 & -0.1001 & -1.0101 & 0.4023 & -0.6062 & 0.1909 & 0.0986 \\    
		&  \multicolumn{1}{|c|}{ (0.0008)} & 0.0192 & 0.0014 & 0.0006 & 0.0050 & 0.0001 & 0.2407 & 0.0392 & 0.0220 & 0.0387 & 0.0047 \\  \hline
		\multicolumn{1}{|l|}{-0.45} &  \multicolumn{1}{|c|}{-0.4560 }  & 1.0521 & 0.3056 & 0.3027 & 0.1711 & -0.1002 & -1.0084 & 0.4014 & -0.6062 & 0.1899 & 0.0994 \\  
		&  \multicolumn{1}{|c|}{ (0.0010) } & 0.0168 & 0.0013 & 0.0005 & 0.0044 & 0.0001 & 0.2130 & 0.0358 & 0.0206 & 0.0401 & 0.0047 \\  \hline
		\multicolumn{1}{|l|}{-0.3} &  \multicolumn{1}{|c|}{-0.3037 } & 1.0455 & 0.3051 & 0.3025  & 0.1745 & -0.1002 & -0.9983 & 0.3966 & -0.6053 & 0.1860 & 0.0988 \\ 
		&  \multicolumn{1}{|c|}{(0.0014)} & 0.0147 & 0.0012 & 0.0005 & 0.0039 & 0.0001 & 0.1991 & 0.0348 & 0.0203 & 0.0443 & 0.0047 \\  \hline
		\multicolumn{1}{|l|}{-0.15} &  \multicolumn{1}{|c|}{-0.1607   } & 1.0404 & 0.3049 & 0.3026 & 0.1769 & -0.1002 & -1.0016 & 0.3973 & -0.6014 & 0.1862 & 0.0991 \\   
		&  \multicolumn{1}{|c|}{ (0.0017) }& 0.0132 & 0.0012 & 0.0005 & 0.0036 & 0.0001 & 0.1775 & 0.0318 & 0.0199 & 0.0485 & 0.0048 \\  \hline
		\multicolumn{1}{|l|}{0} &   \multicolumn{1}{|c|}{0.0093   } & 1.0356 & 0.3050 & 0.3024 & 0.1790 & -0.1002 & -0.9926 & 0.3925 & -0.6007 & 0.1811 & 0.0985 \\   
		&  \multicolumn{1}{|c|}{ (0.0022) } & 0.0111 & 0.0011 & 0.0005 & 0.0032 & 0.0001 & 0.1654 & 0.0312 & 0.0201 & 0.0557 & 0.0048 \\  \hline
		\multicolumn{1}{|l|}{0.15} &  \multicolumn{1}{|c|}{0.1534   } &  1.0324 & 0.3053 & 0.3017 & 0.1802 & -0.1002 & -0.9931 & 0.3924 & -0.6001 & 0.1792 & 0.0984 \\  
		&  \multicolumn{1}{|c|}{ (0.0013)  } & 0.0099 & 0.0012 & 0.0005 & 0.0030 & 0.0001 & 0.1592 & 0.0308 & 0.0205 & 0.0596 & 0.0047 \\  \hline
		\multicolumn{1}{|l|}{0.3} &  \multicolumn{1}{|c|}{ 0.2985  } & 1.0290 & 0.3054 & 0.3013 & 0.1816 & -0.1002 & -0.9876 & 0.3900 & -0.6004 & 0.1789 & 0.0984 \\   
		&  \multicolumn{1}{|c|}{ (0.0013)  } &     0.0089 & 0.0012 & 0.0005 & 0.0028 & 0.0001 & 0.1576 & 0.0315 & 0.0215 & 0.0650 & 0.0046 \\ \hline
		\multicolumn{1}{|l|}{0.45} & \multicolumn{1}{|c|}{0.4492 } &  1.0271 & 0.3067 & 0.3004 & 0.1813 & -0.1004 & -0.9834 & 0.3875 & -0.5990 & 0.1771 & 0.0978 \\   
		&  \multicolumn{1}{|c|}{  (0.0011) }  & 0.0080 & 0.0014 & 0.0005 & 0.0028 & 0.0001 & 0.1521 & 0.0317 & 0.0226 & 0.0720 & 0.0046 \\  \hline
		\multicolumn{1}{|l|}{0.6} &  \multicolumn{1}{|c|}{0.5991  } & 1.0251 & 0.3076 & 0.2994 & 0.1814 & -0.1003 & -0.9770 & 0.3839 & -0.5947 & 0.1745 & 0.0979 \\  
		&  \multicolumn{1}{|c|}{ (0.0008)}  & 0.0073 & 0.0015 & 0.0006 & 0.0028 & 0.0001 & 0.1489 & 0.0333 & 0.0255 & 0.0806 & 0.0046 \\ \hline
		\multicolumn{1}{|l|}{0.75} &  \multicolumn{1}{|c|}{ 0.7493   } & 1.0236 & 0.3100 & 0.2977 & 0.1800 & -0.1003 & -0.9759 & 0.3825 & -0.5921 & 0.1718 & 0.0984 \\   
		&  \multicolumn{1}{|c|}{  (0.0004) } & 0.0070 & 0.0018 & 0.0007 & 0.0032 & 0.0001 & 0.1528 & 0.0351 & 0.0273 & 0.0898 & 0.0046 \\  \hline
		\multicolumn{1}{|l|}{0.9} &  \multicolumn{1}{|c|}{0.9012   } & 1.0237 & 0.3135 & 0.2955 & 0.1767 & -0.1002 & -0.9708 & 0.3798 & -0.5872 & 0.1746 & 0.0991 \\  
		& \multicolumn{1}{|c|}{ (0.0001) } & 0.0066 & 0.0022 & 0.0008 & 0.0038 & 0.0001 & 0.1557 & 0.0370 & 0.0295 & 0.0999 & 0.0047 \\ \hline
	\end{tabular}
	\label{1}
\end{table}
\newpage
\begin{table}[htbp]
	\center 
	\small 
	\caption{Average and MSE for the estimators of the BIP model ($n=500$)}
	\begin{tabular}{|ll|ccccc|ccccc|}
		\cline{3-12}
		\multicolumn{2}{c|}{}  & \multicolumn{5}{c|}{Log INGARCH}  & \multicolumn{5}{c|}{Logit Binary}  \\ \hline
		\multicolumn{2}{|l|}{n = 500} & \multicolumn{1}{c}{$d_1$} & \multicolumn{1}{c}{$A(1,1)$} & \multicolumn{1}{c}{$A(1,2)$} & \multicolumn{1}{c}{$B(1,1)$} & \multicolumn{1}{c|}{$\Gamma(1,1)$} & \multicolumn{1}{c}{$d_2$} & \multicolumn{1}{c}{$A(2,1)$} & \multicolumn{1}{c}{$A(2,2)$} & \multicolumn{1}{c}{$B(2,2)$} & \multicolumn{1}{c|}{$\Gamma(2,1)$} \\ \hline
		\multicolumn{2}{|c|}{$r$}  & 1 & 0.3 & 0.3 & 0.15 & -0.1 & -1 & 0.4 & -0.6 & 0.2 &  0.1 \\ \hline
		\multicolumn{1}{|l|}{-0.9} & \multicolumn{1}{|c|}{-0.9001} & 1.1030 &  0.3193 &   0.3103 &  0.1329 &  -0.1002 & -0.9645 & 0.3868  &  -0.6315 &  0.1896 &  0.1057  \\ 
		& \multicolumn{1}{|c|}{(0.0002)} & 0.0479 &  0.0042 &  0.0017 &  0.0090 &  0.0002 & 0.7927 &  0.1230 &  0.0670 &  0.0707 &  0.0096 \\  \hline
		\multicolumn{1}{|l|}{-0.75} &  \multicolumn{1}{|c|}{ -0.7494 } & 1.0940 &  0.3157 &   0.3080 &  0.1410 &  -0.1002  & -0.9683 &  0.3888 &  -0.6251 &  0.1980 &   0.1061 \\   
		&  \multicolumn{1}{|c|}{ ((0.0008))}  & 0.0410 &  0.0032 &  0.0013  &  0.0076 &  0.0002  & 0.5873 &  0.0923 &  0.0546 &  0.0720 & 0.0095 \\  \hline
		\multicolumn{1}{|l|}{-0.6} &   \multicolumn{1}{|c|}{-0.5955 } & 1.0880 &  0.3134 &   0.3064 &  0.1465 &  -0.1003  & -0.9697 &  0.3886 &  -0.6260 & 0.1932 &   0.1051  \\    
		&  \multicolumn{1}{|c|}{(0.0017)} & 0.0368 &  0.0027 &  0.0011 &  0.0068 & 0.0002  &  0.4670 &  0.0769 &  0.0469 &  0.0753 &  0.0096  \\  \hline
		\multicolumn{1}{|l|}{-0.45} &  \multicolumn{1}{|c|}{-0.4482}  &  1.0810 &  0.3125 &   0.3052 &  0.1507 &  -0.1001  & -0.9730 &  0.3902 &  -0.6239 &  0.1958 &   0.1042 \\  
		&  \multicolumn{1}{|c|}{ (0.0026) } & 0.0322 &  0.0024 &  0.0009 &  0.0060 &  0.0002  & 0.3910 & 0.0655 &  0.0438 &  0.0789 &  0.0097  \\  \hline
		\multicolumn{1}{|l|}{-0.3} &  \multicolumn{1}{|c|}{-0.3059 } & 1.0751 &  0.3121 &   0.3044 &  0.1538 &  -0.1000  & -0.9747 &  0.3913 &  -0.6244 &  0.1958 &   0.1045  \\ 
		&  \multicolumn{1}{|c|}{(0.0025)} & 0.0286 &  0.0022 &  0.0009 &  0.0055 &  0.0002  & 0.3532 &  0.0611 & 0.0413 &  0.0830 &  0.0097  \\  \hline
		\multicolumn{1}{|l|}{-0.15} &  \multicolumn{1}{|c|}{-0.1513} & 1.06798 &  0.3125 &   0.3031 &  0.1568 &  -0.1001 & -0.9826  &  0.3949 &  -0.6205 &  0.1992 &   0.1042  \\   
		&  \multicolumn{1}{|c|}{ (0.0034) } & 0.0253 &  0.0022 &  0.0008 &  0.0052 &  0.0002  & 0.3325 &  0.0583 &  0.0400 &  0.0876 &  0.0098   \\  \hline
		\multicolumn{1}{|l|}{0} &   \multicolumn{1}{|c|}{0.0001} & 1.0644 &  0.3127 &   0.3020 &  0.1584 &  -0.1001  & -0.9851 &  0.3956 & -0.6201 & 0.1979 &   0.1037 \\   
		&  \multicolumn{1}{|c|}{ (0.0032) } & 0.0225 & 0.0023 &  0.0008 &  0.0050 &  0.0002  & 0.3084 &  0.0567 &  0.0400 &  0.0935 &  0.0099  \\  \hline
		\multicolumn{1}{|l|}{0.15} &  \multicolumn{1}{|c|}{0.1519  } &  1.0607 &  0.3140 &   0.3004 &  0.1590 &  -0.1000  & -0.9833 &  0.3947 &  -0.6153 &  0.2024 &   0.1038\\  
		&  \multicolumn{1}{|c|}{ (0.0039)  } & 0.0204 &  0.0024 &  0.0008 &  0.0048 &  0.0002  & 0.3007   & 0.0560 & 0.0413 &  0.1059 &  0.0097 \\  \hline
		\multicolumn{1}{|l|}{0.3} &  \multicolumn{1}{|c|}{0.2978} & 1.0553 &  0.3154 &   0.2989 &  0.1602 &  -0.1000  & -0.9928 & 0.3998 &  -0.6165 &  0.2045 &   0.1033 \\   
		&  \multicolumn{1}{|c|}{(0.0027)} &  0.0186 &  0.0026 &  0.0009 &  0.0048 & 0.0002  & 0.2912 &  0.0557 &  0.0419 &  0.1109 & 0.0095 \\ \hline
		\multicolumn{1}{|l|}{0.45} & \multicolumn{1}{|c|}{0.4550} & 1.0533 &  0.3179 &   0.2971 &  0.1589 &  -0.0998  & -0.9873 &  0.3969 &  -0.6182 &  0.2024 &   0.1020  \\   
		&  \multicolumn{1}{|c|}{(0.0025)}  & 0.0173 & 0.0029 &  0.0010 &  0.0049 &  0.0002  & 0.2945 &  0.0584 &  0.0445 &  0.1205 &  0.0096 \\  \hline
		\multicolumn{1}{|l|}{0.6} &  \multicolumn{1}{|c|}{0.6021} & 1.0524 &  0.3218 &   0.2944 &  0.1559 & -0.0999  & -0.9883 & 0.3979  &  -0.6207 &  0.2031 &  0.1018 \\  
		&  \multicolumn{1}{|c|}{ (0.0015)}  & 0.0161 &  0.0033 &  0.0012 & 0.0050 &  0.0002  & 0.3080 &  0.0622 &  0.04675 &  0.1220 &  0.0095  \\ \hline
		\multicolumn{1}{|l|}{0.75} &  \multicolumn{1}{|c|}{0.7476} & 1.0503 & 0.3263 &   0.2905 &  0.1528 &  -0.0997  & -0.9706 &  0.3866 &  -0.6141 & 0.1863 &   0.1009  \\   
		&  \multicolumn{1}{|c|}{(0.0009)} &  0.0153 &  0.0039 &  0.0014 & 0.0053 &  0.0002 &  0.3362 &  0.0726 & 0.0525 & 0.1386 &  0.0097  \\  \hline
		\multicolumn{1}{|l|}{0.9} &  \multicolumn{1}{|c|}{0.9008} & 1.048 &  0.3318 &   0.2856 &  0.1491 & -0.0996   & -0.9544 &  0.3819 &  -0.6128 &  0.1956 &   0.1004\\  
		& \multicolumn{1}{|c|}{(0.0002)} & 0.0148 &  0.0049 &  0.0017 & 0.0058 &  0.0002  & 0.3423 &  0.0772 &  0.0595 &  0.1504 & 0.0095  \\ \hline
	\end{tabular}
	\label{2}
\end{table}
\newpage
\begin{table}[htbp]
	\center 
	\small 
	\caption{Average and MSE for the estimators of the GAIN model ($n=1000$)}
	\begin{tabular}{|ll|cccc|cccc|}
		\cline{3-10}
		\multicolumn{2}{c|}{}  & \multicolumn{4}{c|}{ GARCH}  & \multicolumn{4}{c|}{INGARCH}  \\ \hline
		\multicolumn{2}{|l|}{n = 1000} & \multicolumn{1}{c}{$d_1$} & \multicolumn{1}{c}{$A(1,1)$} & \multicolumn{1}{c}{$A(1,2)$} & \multicolumn{1}{c|}{$B(1,1)$}   & \multicolumn{1}{c}{$d_2$} & \multicolumn{1}{c}{$A(2,1)$} & \multicolumn{1}{c}{$A(2,2)$} & \multicolumn{1}{c|}{$B(2,2)$}  \\ \hline
		\multicolumn{2}{|c|}{$r$}  & 0.03 &  0.05 &  0.05 &  0.7 &  0.3 &  0.3 &  0.1 &  0.5 \\ \hline
		\multicolumn{1}{|l|}{-0.9} & \multicolumn{1}{|c|}{-0.8995 } &0.0395 & 0.0514  & 0.0534 & 0.6593 & 0.3270  & 0.3088  & 0.0992  & 0.4703  \\ 
		& \multicolumn{1}{|c|}{(0.0001)} & 0.0005 & 0.0008  & 0.0002 & 0.0101 & 0.0094 & 0.0058 & 0.0009 & 0.0133   \\  \hline
		\multicolumn{1}{|l|}{-0.75 } &  \multicolumn{1}{|c|}{-0.7463} & 0.0405  & 0.0498  & 0.0540  & 0.6543 & 0.3291  & 0.3067  & 0.1008  & 0.4670 \\   
		&  \multicolumn{1}{|c|}{ (0.0004)}  & 0.0006 & 0.0008  & 0.0002 & 0.01067 & 0.0098 & 0.0058  & 0.0009 & 0.0142   \\  \hline
		\multicolumn{1}{|l|}{-0.6} &   \multicolumn{1}{|c|}{-0.6006  } & 0.0407  & 0.0504 & 0.0542  & 0.6539 & 0.3304 & 0.3123 & 0.0991 & 0.4669 \\    
		&  \multicolumn{1}{|c|}{ (0.0008)} & 0.0006  & 0.0008 & 0.0002 & 0.0105  & 0.0096  & 0.0058 & 0.0009  & 0.0137  \\  \hline
		\multicolumn{1}{|l|}{-0.45} &  \multicolumn{1}{|c|}{-0.4560}  & 0.0415  & 0.0493  & 0.0537  & 0.6538 & 0.3275 & 0.3094 & 0.1016 & 0.4685  \\  
		&  \multicolumn{1}{|c|}{ (0.0010) } & 0.0006  & 0.0008 & 0.0002 & 0.0102 & 0.0099 & 0.0061  & 0.0009  & 0.0145 \\  \hline
		\multicolumn{1}{|l|}{-0.3} &  \multicolumn{1}{|c|}{-0.3037} &  0.0428 & 0.0514  & 0.0536 & 0.6462  & 0.3252 & 0.3107 & 0.1026 & 0.4693   \\ 
		&  \multicolumn{1}{|c|}{(0.0014)} & 0.0009 & 0.0008 & 0.0002 & 0.0151 & 0.0101 & 0.0063  & 0.0009  & 0.0147 \\  \hline
		\multicolumn{1}{|l|}{-0.15} &  \multicolumn{1}{|c|}{-0.1607} &0.0416 & 0.0526 & 0.0525  & 0.6532  & 0.3340  & 0.3081 & 0.0988 & 0.4646   \\   
		&  \multicolumn{1}{|c|}{ (0.0017) } & 0.0007  & 0.0008  & 0.0002 & 0.0122 & 0.0118  & 0.0057 & 0.0009  & 0.0162  \\  \hline
		\multicolumn{1}{|l|}{0} &   \multicolumn{1}{|c|}{0.0093} & 0.0407 & 0.0523  & 0.0536 & 0.6532  & 0.3357  & 0.3096  & 0.1029 & 0.4578   \\   
		&  \multicolumn{1}{|c|}{ (0.0022) } & 0.0006 & 0.0008  & 0.0002 & 0.0112  & 0.0100 & 0.0058 & 0.0009  & 0.0146  \\  \hline
		\multicolumn{1}{|l|}{0.15} &  \multicolumn{1}{|c|}{0.1534   } & 0.0411  & 0.0515  & 0.0527 & 0.6542 & 0.3295 & 0.3135  & 0.0982  & 0.4692   \\  
		&  \multicolumn{1}{|c|}{ (0.0013)  } & 0.0006 & 0.0008 & 0.0002 & 0.0117 & 0.0109 & 0.0052 &  0.0009 &  0.0147 \\  \hline
		\multicolumn{1}{|l|}{0.3} &  \multicolumn{1}{|c|}{ 0.2985  } & 0.04308 &  0.0518 &  0.0534 &  0.6475 &  0.3266 &  0.3121 &  0.0963 &  0.4729   \\   
		&  \multicolumn{1}{|c|}{ (0.0013)  } & 0.0007 &  0.0008 &  0.0002 &  0.0119 &  0.0107 &  0.0057 &  0.0009 &  0.0154  \\ \hline
		\multicolumn{1}{|l|}{0.45} & \multicolumn{1}{|c|}{0.4492 } & 0.0404 &  0.0511 &  0.0531 & 0.6571 &  0.3371 &  0.3101 &  0.1008 &  0.4591   \\   
		&  \multicolumn{1}{|c|}{  (0.0011) }  &0.0007 &  0.0008 &  0.0002 &  0.0115 & 0.0110 &  0.0059 &  0.0010 &  0.0156 \\  \hline
		\multicolumn{1}{|l|}{0.6} &  \multicolumn{1}{|c|}{0.5991  } & 0.0409 & 0.0526 & 0.0537 & 0.6511  & 0.3247 & 0.3092 & 0.1004 &  0.4712  \\  
		&  \multicolumn{1}{|c|}{ (0.0008)}  & 0.0006 & 0.0008 &  0.0002 &  0.0116 & 0.0088 &  0.0053 &  0.0010 & 0.0120  \\ \hline
		\multicolumn{1}{|l|}{0.75} &  \multicolumn{1}{|c|}{ 0.7493   } & 0.0413 &  0.0514 &  0.0532  & 0.6525 & 0.3378 & 0.3097 &  0.1009 &  0.4574    \\   
		&  \multicolumn{1}{|c|}{  (0.0004) } & 0.0007 &  0.0008 &  0.0002 & 0.0121 &  0.0110 &  0.0052 &  0.0010 &  0.0157 \\  \hline
		\multicolumn{1}{|l|}{0.9} &  \multicolumn{1}{|c|}{0.9012   } & 0.0410 & 0.0530 &  0.0539 & 0.6510 & 0.3214 &  0.3016 &  0.1001 &  0.4777   \\  
		& \multicolumn{1}{|c|}{ (0.0001) } & 0.0005 & 0.0008 & 0.0002 & 0.0098 &   0.0080 &  0.0057 &  0.0010 &  0.0115 \\ \hline
	\end{tabular}
	\label{3}
\end{table}

\newpage
\begin{table}[htbp]
	\center 
	\small 
	\caption{Average and MSE for the estimators of the GAIN model ($n=500$)}
	\begin{tabular}{|ll|cccc|cccc|}
		\cline{3-10}
		\multicolumn{2}{c|}{}  & \multicolumn{4}{c|}{ GARCH}  & \multicolumn{4}{c|}{INGARCH}  \\ \hline
		\multicolumn{2}{|l|}{n = 500} & \multicolumn{1}{c}{$d_1$} & \multicolumn{1}{c}{$A(1,1)$} & \multicolumn{1}{c}{$A(1,2)$} & \multicolumn{1}{c|}{$B(1,1)$}   & \multicolumn{1}{c}{$d_2$} & \multicolumn{1}{c}{$A(2,1)$} & \multicolumn{1}{c}{$A(2,2)$} & \multicolumn{1}{c|}{$B(2,2)$}  \\ \hline
		\multicolumn{2}{|c|}{$r$}  & 0.03 &  0.05 &  0.05 &  0.7 &  0.3 &  0.3 &  0.1 &  0.5 \\ \hline
		\multicolumn{1}{|l|}{-0.9} & \multicolumn{1}{|c|}{-0.8841 } & 0.0497 & 0.0530 &  0.0558 & 0.6173 & 0.3512 &  0.3090 &  0.0974 &  0.4472  \\ 
		& \multicolumn{1}{|c|}{(0.0004)} & 0.0019 &  0.0016 &  0.0004 &  0.0297 & 0.0214 &  0.0125 & 0.0022 &  0.0291  \\  \hline
		\multicolumn{1}{|l|}{-0.75  } &  \multicolumn{1}{|c|}{-0.7213} & 0.0517 &  0.0539 &  0.0575 &  0.6022 &  0.3658 &  0.3204 &  0.1029 &  0.4223\\   
		&  \multicolumn{1}{|c|}{ (0.0013)}  & 0.0019 &  0.0018 &  0.0004 &  0.0346 &  0.0247 &  0.0113 &  0.0021 &  0.0348  \\  \hline
		\multicolumn{1}{|l|}{-0.6} &   \multicolumn{1}{|c|}{-0.5731 } & 0.0539 & 0.0511 &  0.0563 &  0.6045 &   0.3627 & 0.3135 &  0.0980 &  0.4342 \\    
		&  \multicolumn{1}{|c|}{ ( 0.0015)} & 0.0022 &  0.0017 &  0.0004 &  0.0333 &  0.0260 &  0.0118 &  0.0021 &  0.0333 \\  \hline
		\multicolumn{1}{|l|}{-0.45 } &  \multicolumn{1}{|c|}{-0.4238 }  &  0.0514 & 0.0541 &  0.0585 &  0.6040 & 0.3654 & 0.3139 &  0.1026 & 0.4261  \\  
		&  \multicolumn{1}{|c|}{ (0.0019) } & 0.0021 & 0.0017 &  0.0004 &  0.0349 &  0.0260 & 0.0118 &  0.0019 &  0.0335 \\  \hline
		\multicolumn{1}{|l|}{-0.3} &  \multicolumn{1}{|c|}{-0.2828} & 0.0532 &  0.0508 & 0.0564 &  0.6077 &  0.3589 & 0.3139 & 0.1003 &  0.4373  \\ 
		&  \multicolumn{1}{|c|}{(0.0020)} & 0.0022 & 0.0015 & 0.0003 & 0.0321 & 0.0263 &  0.0110 & 0.0020 & 0.0337 \\  \hline
		\multicolumn{1}{|l|}{-0.15} &  \multicolumn{1}{|c|}{-0.1544 } & 0.0529 &  0.0535 & 0.0559 & 0.6058 &  0.35308 &  0.3188 &  0.1003 & 0.4391  \\   
		&  \multicolumn{1}{|c|}{ (0.0017) } & 0.0022 &  0.0018 &  0.0004 &  0.0328 & 0.0231 &  0.0110 &  0.0021 &  0.0323  \\  \hline
		\multicolumn{1}{|l|}{0} &   \multicolumn{1}{|c|}{-0.0030 } & 0.0540 & 0.0571 & 0.0555 &  0.6010 & 0.3611 &  0.3129 &  0.1005 &  0.4351 \\   
		&  \multicolumn{1}{|c|}{ (0.0020) } & 0.0024 &  0.0019 &  0.0004 &  0.0364 & 0.0260 &  0.0126 &  0.0019 &  0.0330  \\  \hline
		\multicolumn{1}{|l|}{0.15} &  \multicolumn{1}{|c|}{0.1469  } & 0.0528 &  0.0508 & 0.0555 & 0.6110 & 0.3733 &  0.3106 &  0.0969 & 0.4263  \\  
		&  \multicolumn{1}{|c|}{ (0.0016)  } & 0.0024 & 0.0015 & 0.0003 & 0.0325 &  0.0309 &  0.0105 &  0.0019 &  0.0393  \\  \hline
		\multicolumn{1}{|l|}{0.3} &  \multicolumn{1}{|c|}{0.2866} & 0.0531 & 0.0486 & 0.0543 & 0.6135 &  0.3479 &  0.3148 &  0.0964 &  0.4477 \\   
		&  \multicolumn{1}{|c|}{ (0.0017)  } & 0.0023 & 0.0016 & 0.0003 &  0.0331 &  0.0237 &  0.0117 & 0.0018 &  0.0308  \\ \hline
		\multicolumn{1}{|l|}{0.45} & \multicolumn{1}{|c|}{0.4233 } & 0.0533 &  0.0548 &  0.0559 &  0.6032 &  0.3641 &  0.3186 &  0.0985 & 0.4288   \\   
		&  \multicolumn{1}{|c|}{  (0.0019) }  & 0.0025 &  0.0018 &  0.0004 &  0.0366 &  0.0261 & 0.0117 &  0.0019 & 0.0353 \\  \hline
		\multicolumn{1}{|l|}{0.6} &  \multicolumn{1}{|c|}{0.5699} & 0.0489 &  0.0548 & 0.0567 & 0.6149 &  0.3537 & 0.3099 &  0.0988 &  0.4442  \\  
		&  \multicolumn{1}{|c|}{ (0.0017)}  & 0.0017 &  0.0017 &  0.0004 &  0.0285 &  0.0222 &  0.0113 &  0.0019 &  0.0296   \\ \hline
		\multicolumn{1}{|l|}{0.75} &  \multicolumn{1}{|c|}{0.7210} &0.0520 &  0.0551 &  0.0557 &  0.6075 &  0.3692 &  0.3188 &  0.1014 &  0.4226    \\   
		&  \multicolumn{1}{|c|}{  (0.0012) } & 0.0022 & 0.0018 &  0.0004 &  0.0341 & 0.0257 &  0.0115 &  0.0023 &  0.0344  \\  \hline
		\multicolumn{1}{|l|}{0.9} &  \multicolumn{1}{|c|}{0.8835 } &0.0509 & 0.0568 & 0.0554 &  0.6099 &  0.3489 &  0.3168 &  0.0997 & 0.4444  \\  
		& \multicolumn{1}{|c|}{ (0.0004) } & 0.0018 &  0.0020 &  0.0004 &  0.0301 &  0.0205 &  0.0136 &  0.0019 & 0.0294 \\ \hline
	\end{tabular}
	\label{4}
\end{table}

\bibliographystyle{plainnat}
\bibliography{biblio}
\end{document}